\documentclass[pra,showpacs,superscriptaddress,amssymb,twocolumn,longbibliography,notitlepage]{revtex4-1}
\usepackage{graphicx}
\usepackage{amsmath}
\usepackage{amsthm}
\usepackage{amssymb}
\usepackage[usenames]{color}
\usepackage{hyperref}
\usepackage{listings}
\usepackage{cancel}
\usepackage[inline]{enumitem}

\hypersetup{
    colorlinks=true,       
    linkcolor=cyan,          
    citecolor=magenta,        
    filecolor=magenta,      
    urlcolor=blue,           
    runcolor=cyan
}

\usepackage{bm}
\usepackage{threeparttable}
\usepackage{subfigure}
\usepackage{upgreek }
\usepackage{marginnote}
\usepackage{braket}
\newcommand{\beq}{\begin{equation}}
\newcommand{\eeq}{\end{equation}}
\newcommand{\beqnn}{\begin{equation*}}
\newcommand{\eeqnn}{\end{equation*}}
\newcommand{\bea}{\begin{eqnarray}}
\newcommand{\eea}{\end{eqnarray}}
\newcommand{\beann}{\begin{eqnarray*}}
\newcommand{\eeann}{\end{eqnarray*}}
\newcommand{\bes} {\begin{subequations}}
\newcommand{\ees} {\end{subequations}}

\newcommand{\Tr}{\mathrm{Tr}}

\newcommand{\abs}[1]{\ensuremath{\left| #1 \right|}}

\def\Tr{\operatorname{Tr}}

\def\pr{\prime}

\def\col{\text{col}}

\begin{document}
\title{Sensitivity of quantum speedup by quantum annealing to a noisy oracle}


\author{Siddharth Muthukrishnan}\email{muthukri@usc.edu}
\affiliation{Department of Physics and Astronomy, University of Southern California, Los Angeles, California 90089, USA}
\affiliation{Center for Quantum Information Science \& Technology, University of Southern California, Los Angeles, California 90089, USA}

\author{Tameem Albash}
\affiliation{Department of Physics and Astronomy, University of Southern California, Los Angeles, California 90089, USA}
\affiliation{Center for Quantum Information Science \& Technology, University of Southern California, Los Angeles, California 90089, USA}
\affiliation{Information Sciences Institute, University of Southern California, Marina Del Rey, California, USA}

\author{Daniel A. Lidar}
\affiliation{Department of Physics and Astronomy, University of Southern California, Los Angeles, California 90089, USA}
\affiliation{Center for Quantum Information Science \& Technology, University of Southern California, Los Angeles, California 90089, USA}
\affiliation{Department of Electrical and Computer Engineering, University of Southern California, Los Angeles, California 90089, USA}
\affiliation{Department of Chemistry, University of Southern California, Los Angeles, California 90089, USA}

\begin{abstract}
The glued-trees problem is the only example known to date for which quantum annealing provides an exponential speedup, albeit by partly using excited state evolution, in an oracular setting. How robust is this speedup to noise on the oracle? To answer this, we construct phenomenological short-range and long-range noise models, and noise models that break or preserve the reflection symmetry of the spectrum. We show that under the long-range noise models an exponential quantum speedup is retained.
However, we argue that a classical algorithm with an equivalent long-range noise model also exhibits an exponential speedup over the noiseless model. In the quantum setting the long-range noise is able to lift the spectral gap of the problem so that the evolution changes from  diabatic to adiabatic. In the classical setting, long-range noise creates a significant probability of the walker landing directly on the EXIT vertex. Under short-range noise the exponential speedup is lost, but a polynomial quantum speedup is retained for sufficiently weak noise. In contrast to noise range, we find that breaking of spectral symmetry by the noise has no significant impact on the performance of the noisy algorithms. Our results about the long-range models highlight that care must be taken in selecting phenomenological noise models so as not to change the nature of the computational problem. We conclude from the short-range noise model results that the exponential speedup in the glued-trees problem is not robust to noise, but a polynomial quantum speedup is still possible. 
\end{abstract}

\maketitle

\section{Introduction}

Quantum annealing (QA) \cite{kadowaki_quantum_1998,farhi_quantum_2001,finnila_quantum_1994,Brooke1999,Santoro} usually refers
to a family of analog quantum optimization algorithms that interpolate between an initial Hamiltonian whose ground state is easy to prepare and a final Hamiltonian whose ground state is the answer to the optimization problem we want to solve~\cite{Albash-Lidar:RMP}. 
Typically, QA is operated adiabatically, which means that the interpolation timescale $t_f$ (also referred to as the annealing time) is much larger than the smallest energy gap between the ground state and the first excited state that is encountered along the interpolation. 
The adiabatic theorem for closed system dynamics provides a guarantee that for a sufficiently long $t_f$, the evolution reaches the ground state of the final Hamiltonian with high probability (see, e.g., Ref.~\cite{Jansen:07} for a rigorous statement). 

We can also consider QA operated non-adiabatically. Here too, the goal is to end the evolution with the system in the ground state of the final Hamiltonian, but the system can undergo diabatic transitions to excited states and return to the ground state. To further complicate matters, QA can also refer to a version of open system analog quantum optimization algorithms operating at non-zero temperature~\cite{RevModPhys.80.1061}.

In this work, we consider a particular diabatic, oracular QA algorithm for solving the glued-trees problem, which we modify by the addition of noise to the oracle. 
The glued-trees problem was first introduced in Ref.~\cite{childs2003exponential}, where it was shown that any classical algorithm must necessarily take exponential time to solve this problem, and a quantum walk algorithm was presented which solves the problem in polynomial time. Subsequently, a diabatic QA algorithm was presented which also solves the problem in polynomial time~\cite{Somma:2012kx}. This is so far the only explicit QA algorithm for which an exponential speedup is known. The QA evolution in the algorithm takes the system from the ground state to the first excited state, then back down to the ground state. This transition from and back to the ground state is enabled by the Hamiltonian spectrum, which
is symmetric about the middle of evolution. 

Oracular models may not be practical examples of quantum speedups because it is highly non-trivial to construct an oracle in a way that does not assume that we already know the answer to the problem at hand; and, even if we could do so, oracular Hamiltonians acting on $n$ spins typically involve $n$-body operators. However, they provide insights into the mechanisms and boundaries of quantum speedups, and can sometimes serve as stepping stones to more practical, non-oracular algorithms \cite{Simon:94,Shor:94}. In this work, we address the question of whether the exponential speedup of the QA glued-trees algorithm is robust under noise. The noise models we consider are phenomenological and add a time-independent random matrix with Gaussian entries to the interpolating Hamiltonian. Such noise is more appropriately viewed as a model of control errors than as originating from a system-bath interaction~\cite{Breuer:2002}. We consider two dichotomies of noise models. One dichotomy is between noise models which induce long-range interactions among distant nodes in the graph and noise models which only induce interactions between nearest-neighbor nodes. The other dichotomy is between noise models which break a certain reflection symmetry in the spectrum and noise models which preserve the reflection symmetry.

Our noise models are motivated by three concerns. First, they offer ways to perturb features of the problem that are considered explanatorily relevant to the performance of the QA algorithm. This will become clear later, but the main idea can be illustrated as follows. The QA algorithm described in Ref.~\cite{Somma:2012kx} works reliably because the spectrum is symmetric upon reflection about the middle of the evolution. This symmetry guarantees that if the system is excited to the first excited state in the first half of the evolution due to the presence of an exponentially small energy gap, the system will then encounter the same exponentially small gap in the second half of the evolution and return back to the ground state. Therefore, a perturbation that breaks this reflection symmetry offers a control knob to explore the importance of this symmetry.  Second, given that this is an oracle problem, in order to obtain physical noise models, we need to consider physical realizations of the oracle. But oracles are generically unrealizable as local Hamiltonians. Thus, in the absence of physical implementations, we assume the noise is Gaussian at the oracle level. Finally, we choose these noise models because they allow for a numerical and analytical treatment to reasonably large system sizes.

We now summarize our results. We find that for the long-range noise models, the quantum dynamics show an exponential  speedup over classical algorithms that respect the glued-trees graph-structure. However, this speedup is misleading because an exponential speedup is also observed for a \emph{classical} algorithm with long-range transition terms. More precisely, the long-range noise corresponds to a classical random walk on a graph containing edges connecting \emph{any} two columns (see Fig.~\ref{fig:gt}), which allows for a sufficiently high probability for the random walker to jump directly to the EXIT vertex. Meanwhile, we find that the quantum dynamics with the long-range noise exhibit a speedup because of a perturbative lifting of the spectral gap, which turns dynamics that were diabatic in the noiseless setting to dynamics that are adiabatic in the noisy setting. We also observe that the short-range noise models lose the exponential quantum speedup over the noiseless classical algorithm, but they do show a \emph{polynomial} speedup for sufficiently small values of the noise strength~\footnote{An algorithm $A$ has an exponential (polynomial) speedup over another algorithm $B$ if the asymptotic scaling of $A$ is an exponential (polynomial) function of the asymptotic scaling of $B$.}.

The paper is organized as follows. In Sec.~\ref{sec:gtproblem}, we describe the glued-trees problem and the QA algorithm that solves it. In Sec.~\ref{sec:noisemodels}, we describe the noise models that we study. In Sec.~\ref{sec:numericalresults}, we present numerical results on how the performance of the algorithm changes under the different noise models. In Sec.~\ref{sec:explanation}, we provide an explanation for these results and we conclude in Sec.~\ref{sec:conclusion}.

\section{The Glued-Trees problem}
\label{sec:gtproblem}

\begin{figure}[t]
\centering
\includegraphics[width=\columnwidth,height=0.6\columnwidth]{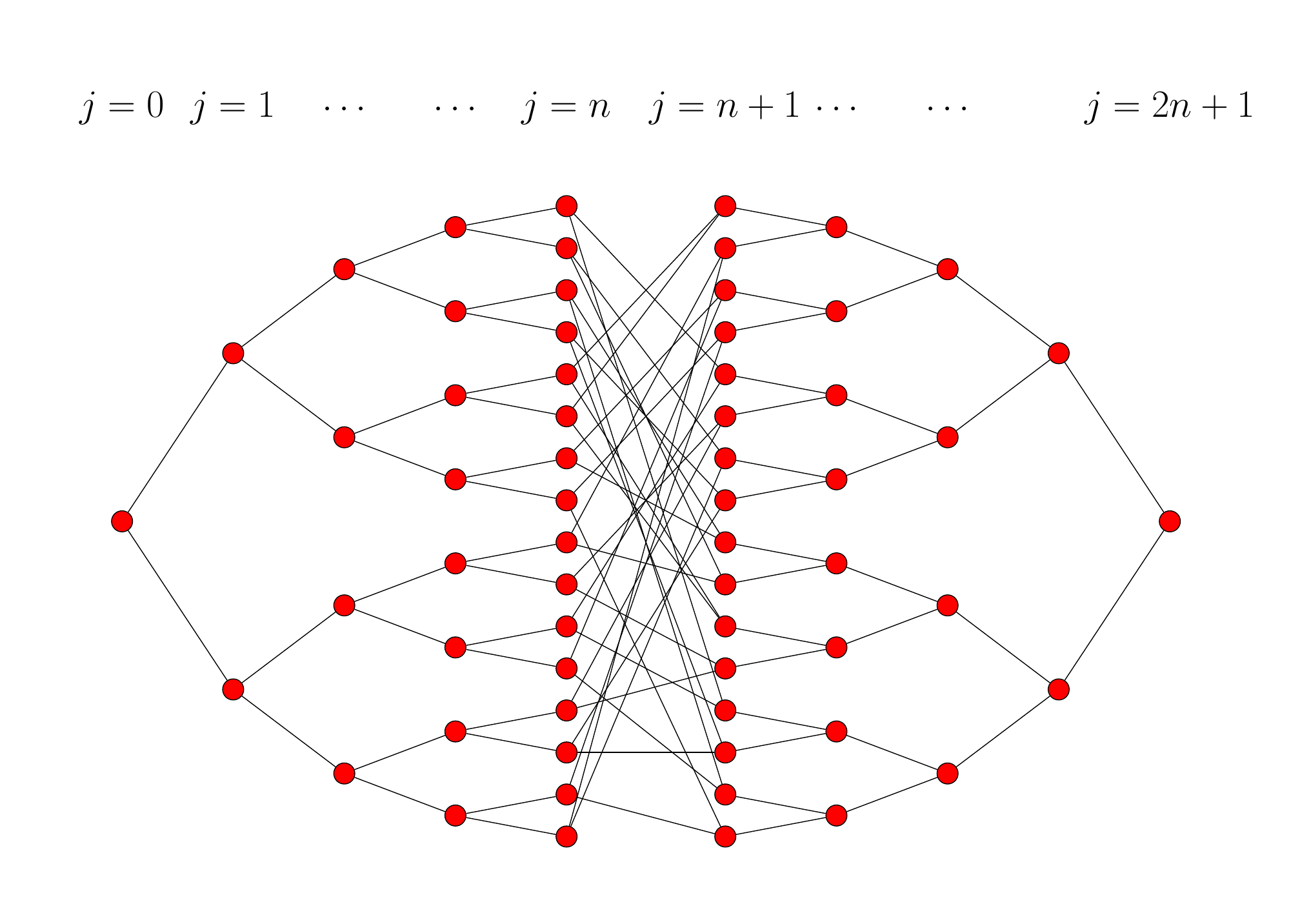}
\caption{(Color online) The graph structure of the glued trees problem. $j=0,1,2,\dots,n,n+1,\dots,2n+1$ indexes the different columns of the graph starting from one vertex to another. }
\label{fig:gt}
\end{figure}

Consider two identical perfect binary trees, of depth $n$, glued together as depicted in Fig.~\ref{fig:gt}. The gluing is done such that each leaf on one tree is randomly joined to two leaves on the other tree, and vice versa. This ensures that every vertex in the graph, except the two root vertices, have degree $3$. One root node is called the ENTRANCE vertex and the other root node is called the EXIT vertex. Starting from the ENTRANCE vertex, the objective is to find the EXIT vertex~\footnote{That all the vertices, except the ENTRANCE and the EXIT vertices, have equal degree is crucial to avoid the easy solution of this problem by a backtracking classical random walk. See Ref.~\cite{childs2003exponential}.}.

All the vertices have labels. A classical algorithm can query the oracle with the label of any vertex, and the oracle returns the labels of the vertices connected to the given vertex. A quantum algorithm can query the oracle with the label of any vertex and the oracle will return a uniform superposition over all the vertices connected to it. Thus, the oracle encodes the adjacency matrix $A$ of the glued-trees graph. Since $n$ is the depth of one of the binary trees, the total number of vertices in the glued trees is $2^{n+2}-2 = \mathcal{O}(2^n)$, which is the minimum number of distinct labels we need. Therefore, the entire graph can be labeled using $(n+2)$-length bitstrings. But this labeling system is insufficient to make the problem hard for classical algorithms. Rather, to prove classical hardness, there need to be exponentially more labels than vertices \cite{childs2003exponential}.  It turns out to be sufficient to choose labels as randomly chosen from the set of $2n$-length bitstrings.

Under this labeling scheme, Ref.~\cite{childs2003exponential} showed that any classical algorithm that makes fewer than $2^{n/6}$ queries to the oracle, will not be able to find the EXIT vertex with probability greater than $4\times 2^{-n/6}$. This means that it will at least take a time $\Omega(2^{n/3})$ to find the EXIT vertex, because in order to boost the success probability we must repeat the algorithm $2^{n/6}$ times. 

On the other hand, it was shown in Ref.~\cite{childs2003exponential} that a quantum walk algorithm which starts from the ENTRANCE vertex and evolves under the Hamiltonian equal to the adjacency matrix of the graph, can find the EXIT vertex with probability $\mathcal{O}(\frac{1}{n})$ if the algorithm is run for times chosen uniformly at random in the interval $[0, \mathcal{O}(n^4)]$. This means we can can get a probability of success arbitrarily close to $1$ by simply repeating the algorithm $\mathcal{O}(n)$ times, and therefore the algorithm will take at most $\mathcal{O}(n^5)$ time. This yields an exponential speedup over the classical algorithm~\footnote{The labeling scheme used for the vertices does not affect the performance of the quantum algorithm.}.

\subsection{The quantum annealing algorithm}

We now turn to the QA algorithm for the glued-trees problem presented in Ref.~\cite{Somma:2012kx}. The initial Hamiltonian is taken to be the projector onto the ENTRANCE vertex: $H_0 = -\ket{\mathrm{ENTRANCE}}\bra{\mathrm{ENTRANCE}}$, such that the initial state of the system coincides with the ground state of the Hamiltonian. The final Hamiltonian is the projector onto the EXIT vertex: $H_1 = -\ket{\mathrm{EXIT}}\bra{\mathrm{EXIT}}$. We then interpolate between these projectors while turning on and off the adjacency matrix $A$:
\beq \label{eq:QAHam}
H(s) = (1-s) \alpha H_0 - s(1-s) A + s \alpha H_1,
\eeq
with $s = t/{t_f}\in [0,1]$, where $t$ is the physical time and $t_f$ is the total evolution time. Also, $0 < \alpha < \frac{1}{2}$ is a constant. We set $\hbar = 1$ throughout. In Ref.~\cite{Somma:2012kx}, it was shown that if $t_f = \mathcal{O}(n^6)$, then the above interpolation ends with sufficiently high probability in the ground state of $H_1$, the EXIT vertex.

With the initial state being $\ket{\mathrm{ENTRANCE}}$, the evolution associated with the Hamiltonian in Eq.~\eqref{eq:QAHam} confines the system to the subspace spanned by the \emph{column basis}, whose elements are defined as
\beq \label{eq:colbasisdef}
\ket{\col_j} \equiv \frac{1}{\sqrt{N_j}} \sum_{a \in \text{column } j} \ket{a},
\eeq
where $\ket{a}$ denotes the state associated with a vertex in column $j$ with label $a$ and
\beq
N_j = \begin{cases} 2^j \ , &  0 \leq j \leq n \\
				2^{2n+1-j} \ , & n+1 \leq j \leq 2n+1
	\end{cases}
\eeq
is the number of vertices in column $j$ (there are $2n+2$ columns in total). It is straightforward to show (see Appendix~\ref{app:colbasis}) that in the column basis, the matrix elements of the Hamiltonian [Eq.~\eqref{eq:QAHam}] are
\bes \label{eq:colbasisH}
\begin{align} 
H_{0,0} &= -\alpha (1-s) \label{eq:colENT} \\ 
H_{j,j+1} = H_{j+1,j} &= -s(1-s) \text{  for  }  j \neq n  \label{eq:coladj}\\
H_{n,n+1} = H_{n+1,n} &= -\sqrt{2} s (1-s) \label{eq:colglue} \\
H_{2n+1,2n+1} &= -\alpha s. \label{eq:colEXIT}
\end{align}
\ees

\subsubsection{Reflection symmetry}

This Hamiltonian is invariant under the composition of two transformations, which together we call the \emph{reflection symmetry}. The first transformation is the reflection of the graph around the central glue. In the column basis, this is represented by the permutation matrix $P$ which has $1$'s on the anti-diagonal and $0$'s everywhere else,
\beq
P_{ij} = \delta_{i,2n+1-j}, \quad i,j \in \{0,1,2,\dots,(2n+1)\}.
\eeq
The second transformation is $s \mapsto (1-s)$: the reflection of the interpolation parameter $s$ around $s=0.5$.
The reflection symmetry is the invariance of the Hamiltonian [Eq.~\eqref{eq:QAHam}] under the composition of these two transformations:
\beq \label{eq:refsymm}
H(s) = PH(1-s)P. 
\eeq

One consequence of the reflection symmetry is that the spectrum of the Hamiltonian is symmetric under the second transformation $s \mapsto (1-s)$ alone. This is because Eq.~\eqref{eq:refsymm} implies that $s \mapsto (1-s)$ corresponds to effectively conjugating the Hamiltonian by $P$, and since $P$ is unitary, the spectrum is unchanged. Therefore,
\beq \label{eq:eigvalsym}
E_k(s) = E_k(1-s) \text{   for   } k \in \{0,1,2,\dots,(2n+1)\}.
\eeq

Another consequence of the symmetry is that if $\ket{\phi_k(s)}$ is the $k$-th eigenstate of $H(s)$, then
\bes
\begin{align}
H(s)\ket{\phi_k(s)} &= E_k(s) \ket{\phi_k(s)} \\
\implies PH(s)P^\dagger P\ket{\phi_k(s)} &= E_k(s) P \ket{\phi_k(s)} \\
\implies H(1-s) (P\ket{\phi_k(s)}) &= E_k(s) (P \ket{\phi_k(s)}) \\
\implies H(1-s) (P\ket{\phi_k(s)}) &= E_k(1-s) (P \ket{\phi_k(s)}) \\
\implies \ket{\phi_k(1-s)} &= P\ket{\phi_k(s)}. \label{eq:eigvecsym}
\end{align}
\ees
Together, Eqs.~\eqref{eq:eigvalsym} and~\eqref{eq:eigvecsym} imply that $H(1-s)$ has the same eigenvalues as $H(s)$ and that the eigenvectors of $H(1-s)$ are the reversed-in-column-basis eigenvectors of $H(s)$.

\subsection{Dynamics}\label{sec:gtqadynamics}

\begin{figure}[!htbp]
\centering
\includegraphics[width=\columnwidth]{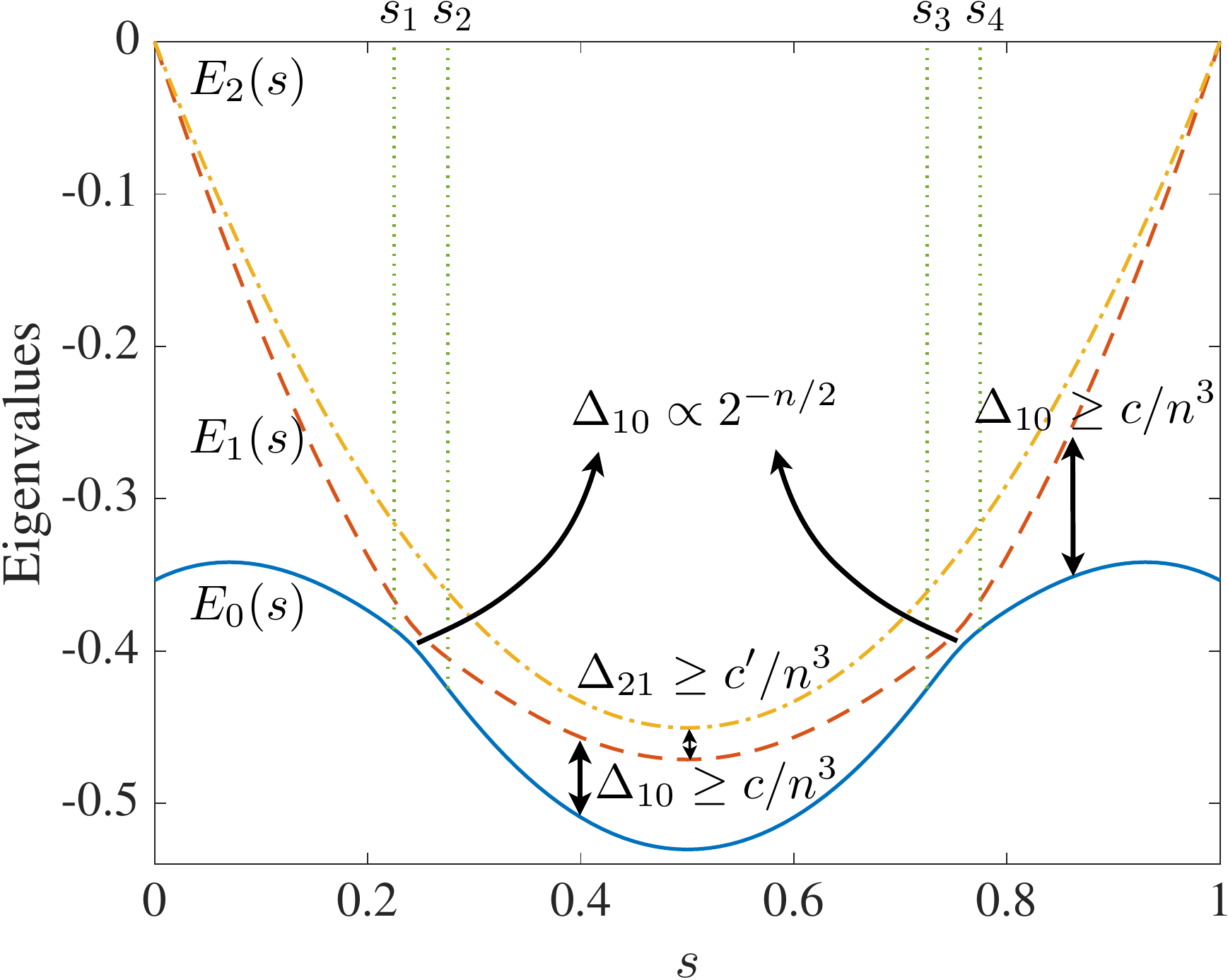}
\caption{(Color online) The smallest three eigenvalues of the Hamiltonian for the case of $n=6$ and $\alpha = 1/\sqrt{8}$. We choose a small $n$ so that the exponentially small gaps are visible. (This image from Ref.~\cite{Albash-Lidar:RMP}.)}
\label{fig:spectrum}
\end{figure}

As shown in Ref.~\cite{Somma:2012kx}, the key features of the Hamiltonian that results in a polynomial time performance are the scalings of the avoided level-crossings in the spectrum, depicted in Fig.~\ref{fig:spectrum}. The evolution that solves the problem in polynomial time is as follows. The system starts in the ground state of the Hamiltonian at $s=0$ (i.e., the ENTRANCE vertex). In the optimal evolution, the system diabatically transitions to the first excited state at the first exponentially small gap (between $s_1$ and $s_2$). Then, it adiabatically follows the first excited state and does not transition to the second excited state because of the polynomially large gap between the first and second excited states. Finally, the system returns back down to ground state through the second exponentially small gap (between $s_3$ and $s_4$). At the end of the annealing evolution described above, we get the EXIT vertex with high probability, as long as the evolution time $t_f$ is chosen to scale as $\mathcal{O}(n^6)$. 

\begin{figure}[!htbp]
\centering
\includegraphics[width=\columnwidth]{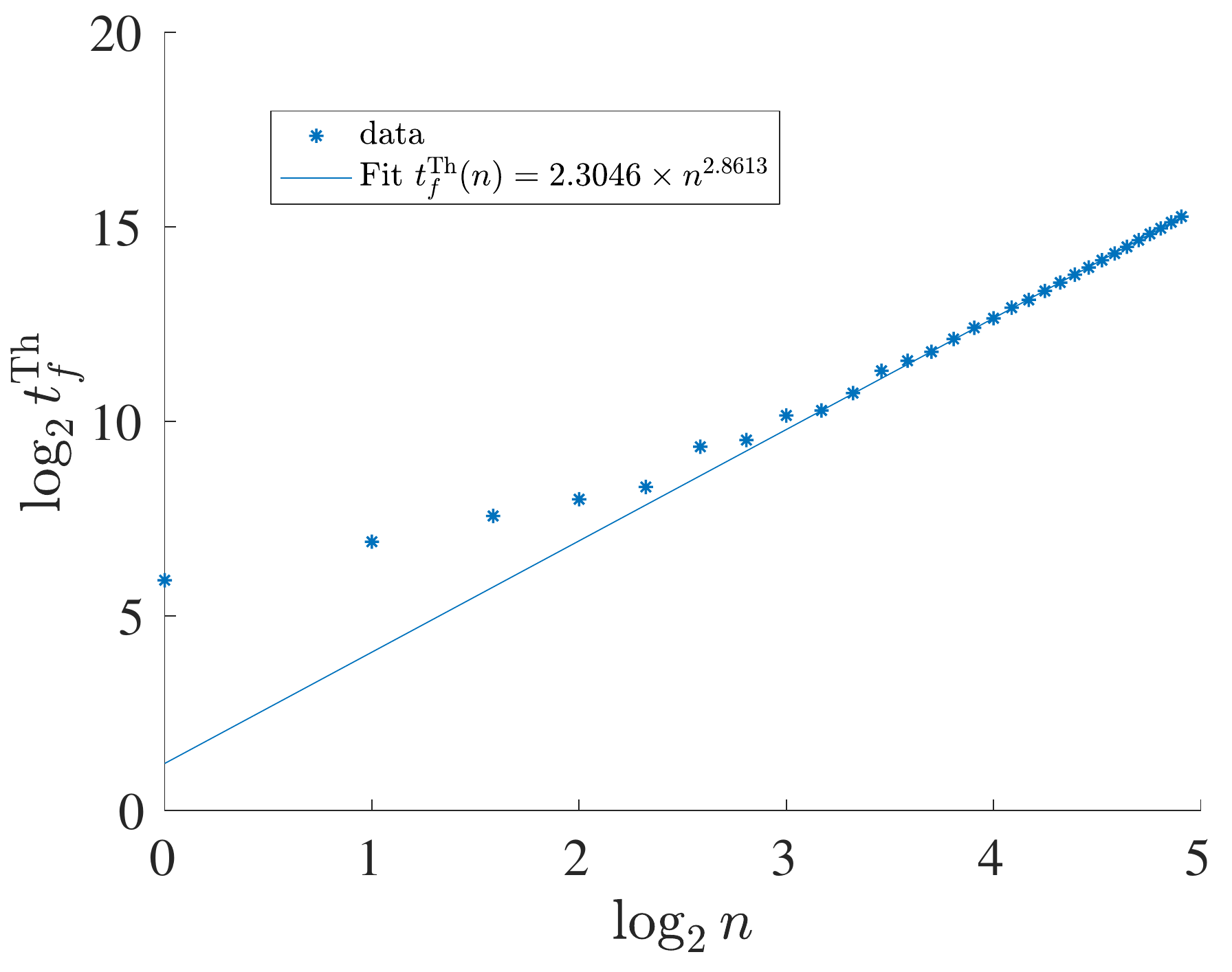}
\caption{The minimum time required to reach a success probability of $p_\mathrm{Th} = 0.95$ as a function of size size $n$ for the noiseless quantum annealing glued trees algorithm. The solid line corresponds to a scaling of $n^{2.8613}$.}
\label{fig:gtnoiselessqa}
\end{figure}

Since the scaling $\mathcal{O}(n^6)$ is an analytically derived upper bound, we expect and find the scaling obtained via numerical simulations to be better. To see this, let us define the threshold annealing time to be the minimum time required for the success probability (where success is defined as reaching the EXIT vertex) to reach a threshold probability $p_\mathrm{Th}$:
\beq \label{eqt:tf}
t_f^\mathrm{Th}(n) \equiv \min \{t_f : p_\mathrm{GS}(t_f) \geq p_\mathrm{Th} \},
\eeq
(henceforth, we choose $p_\mathrm{Th} = 0.95$). In Fig.~\ref{fig:gtnoiselessqa}, we plot the scaling of $t_f^\mathrm{Th}(n)$ for the QA algorithm for the glued trees problem. The scaling is $\mathcal{O}(n^{2.86})$, which is significantly faster than $\mathcal{O}(n^6)$.

\begin{figure}[t]
\centering
\includegraphics[width=\columnwidth]{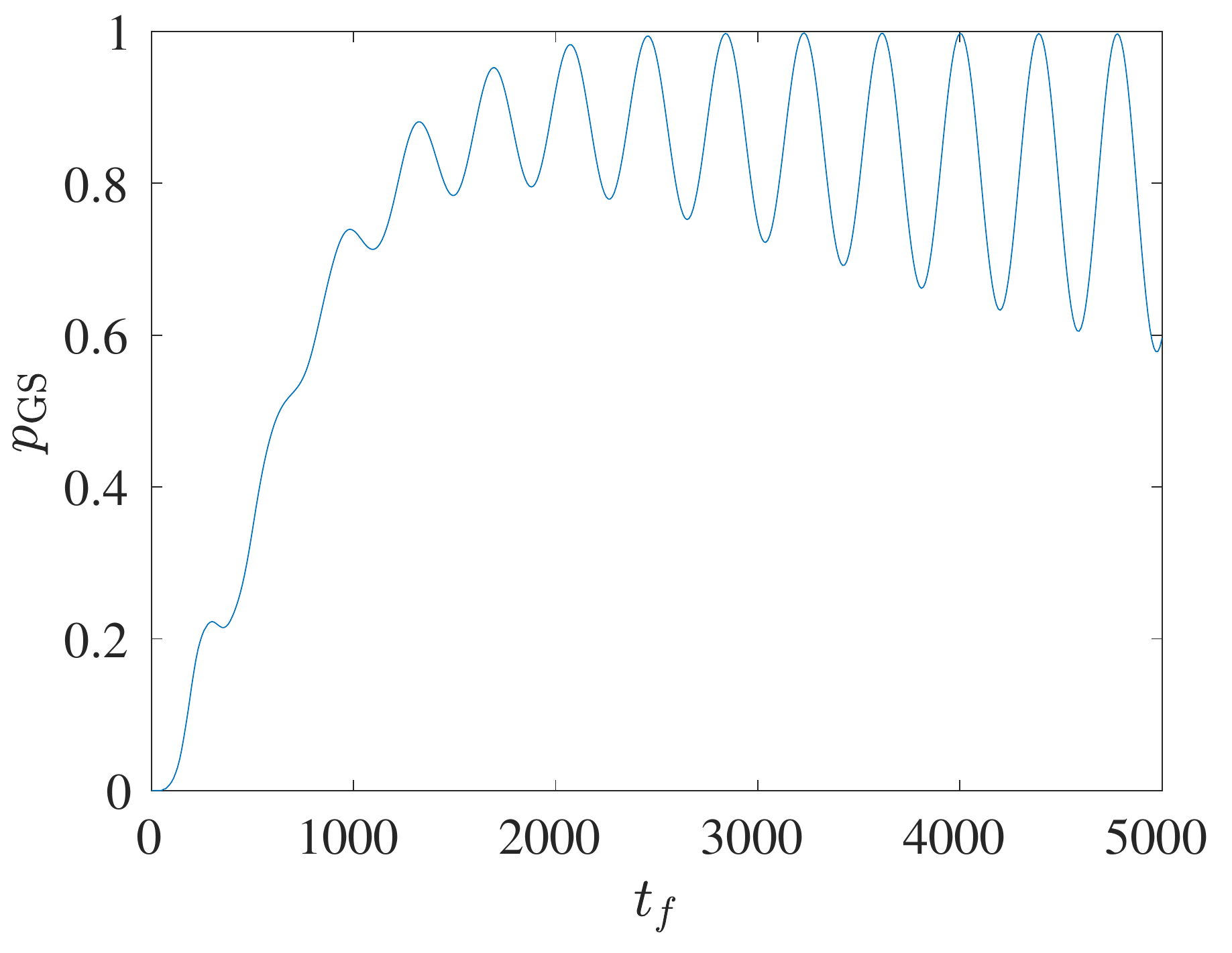}
\caption{Probability of finding the ground state at the end of evolution as a function of $t_f$ for the noiseless glued-trees quantum anneal at problem size $n=10$. Strong oscillations are observed, indicating that the optimal time at which we should terminate the algorithm is sub-adiabatic.}
\label{fig:pgstfqanoiseless}
\end{figure}

It is instructive to examine the $p_\mathrm{GS}(t_f)$ function. This is exhibited for the case $n=10$ in Fig.~\ref{fig:pgstfqanoiseless}. For $n=10$, the threshold timescale is $t_f^\mathrm{Th}(10) = 1690$.
This corresponds to the second peak in the oscillations. In general, the QA algorithm works by being in a region of the $p_\mathrm{GS}(t_f)$ function before adiabaticity is achieved. 

\begin{figure*}[t]
\subfigure[]{\includegraphics[width = \columnwidth]{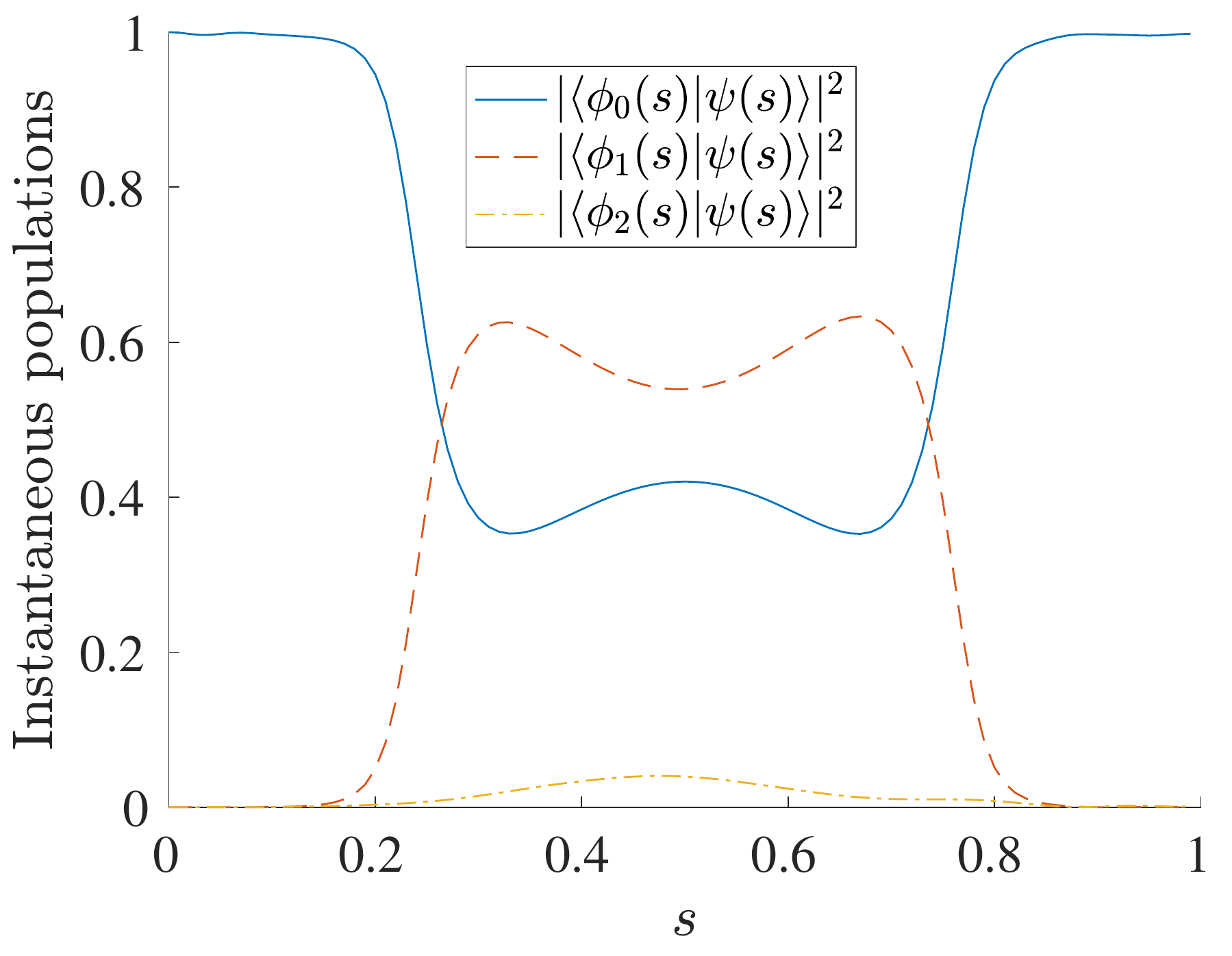}\label{fig:gtqatf250}}
\subfigure[]{\includegraphics[width = \columnwidth]{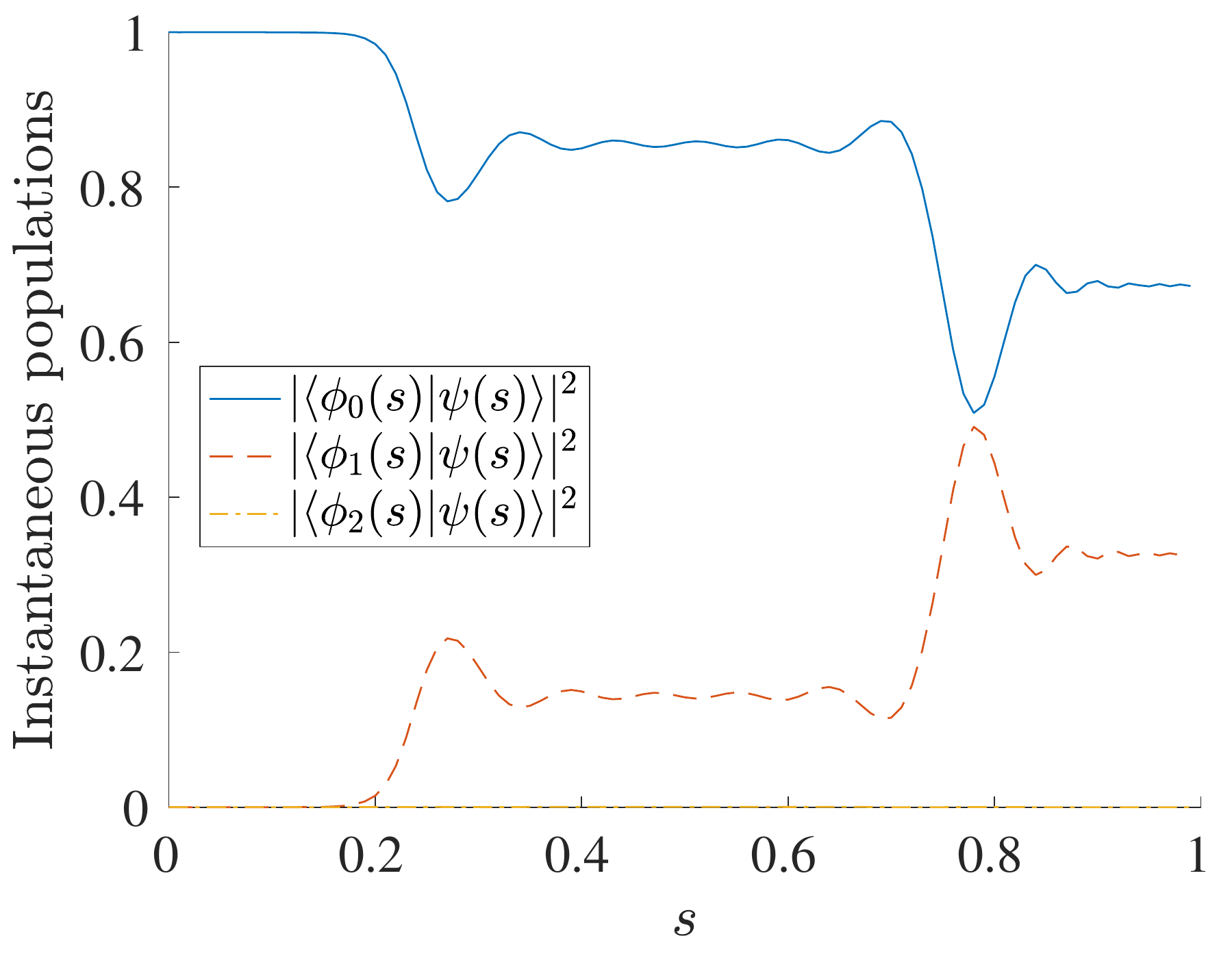} \label{fig:gtqatf1000}}
\subfigure[]{\includegraphics[width = \columnwidth]{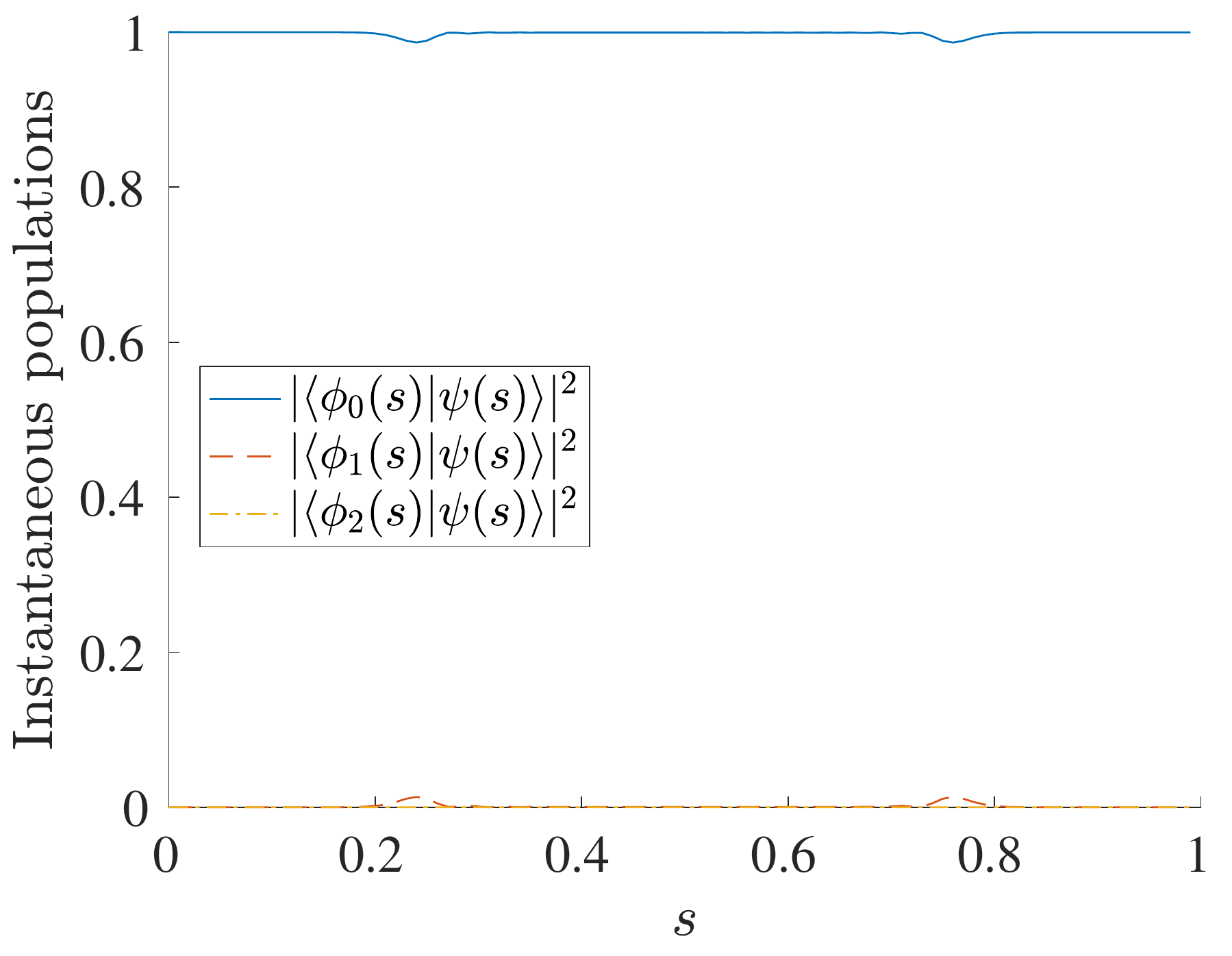}\label{fig:gtqatf4000}}
\subfigure[]{\includegraphics[width=\columnwidth]{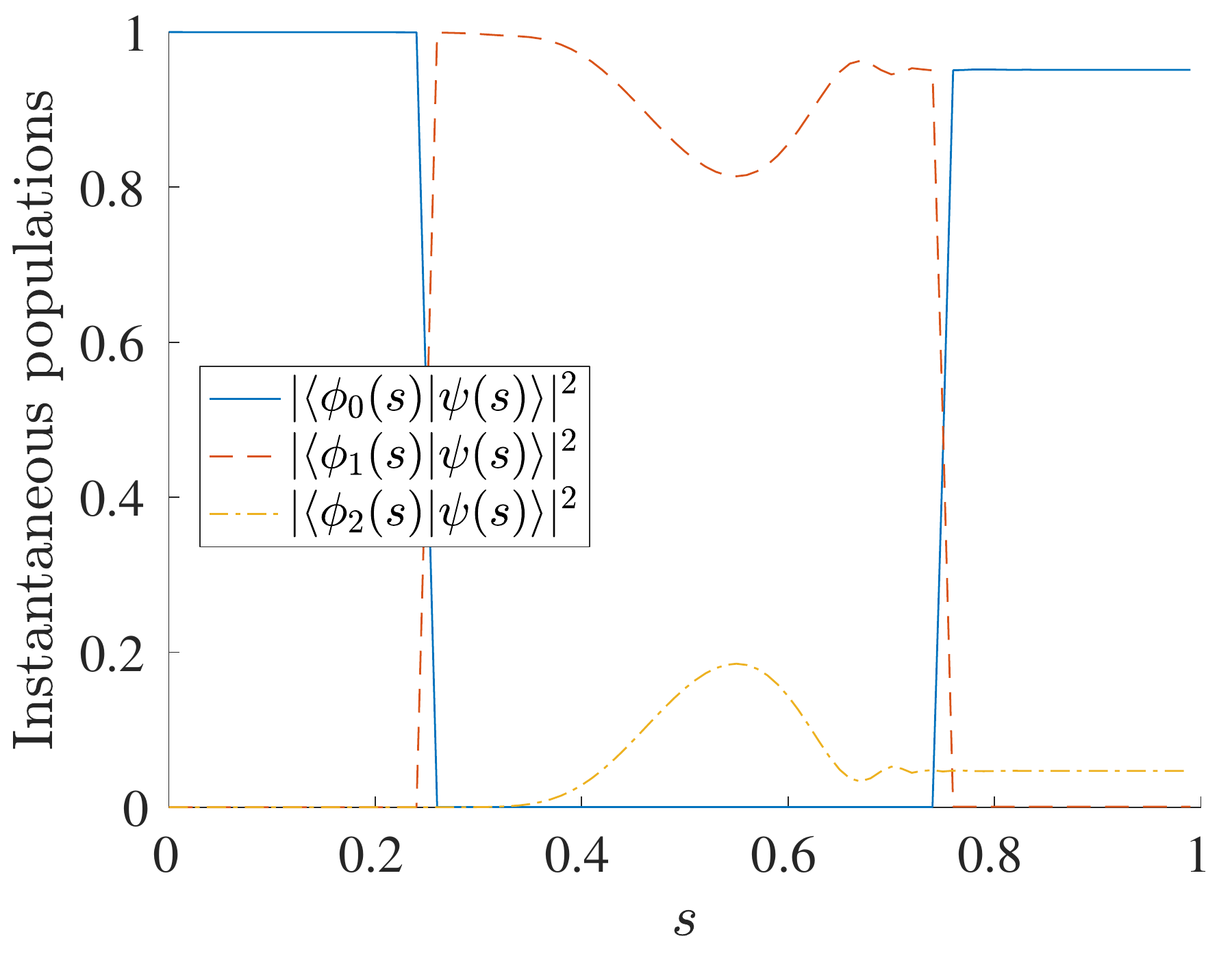}\label{fig:noiselesspopsn20}}
\caption{(Color online) Populations in the instantaneous ground state, first excited state, and second excited state as a function of the anneal parameter $s$ for the glued-trees problem without any added noise. (a) $n=4$, $t_f = 268$; (b): $n=4$, $t_f = 1000$; (c): $n=4$, $t_f = 4000$; (d) $n=20$ at $t_f = t_f^\mathrm{Th}(20) = 12125$ [see Eq.~\eqref{eqt:tf}].
}
\label{fig:gspopn4}
\end{figure*}

It is also instructive to examine what the dynamics look like at different evolution timescales. We examine the populations in the instantaneous ground state, first excited state, and the second excited state as a function of the interpolation parameter $s$ for $n=4,20$ in Fig.~\ref{fig:gspopn4}. For $n=4$, at relatively small annealing times the evolution is close to optimal: the population starts off in the ground state, enters the first excited state at the first exponentially small gap, and returns to the ground state at the second exponentially small gap. At longer annealing times the dynamics is closer to adiabatic, with some interesting fluctuations that arise around the exponentially small gaps. For $n=20$, at the threshold annealing time, the evolution is optimal and exhibits sharp transitions.

\begin{figure*}[t]
\subfigure[]{\includegraphics[width = \columnwidth]{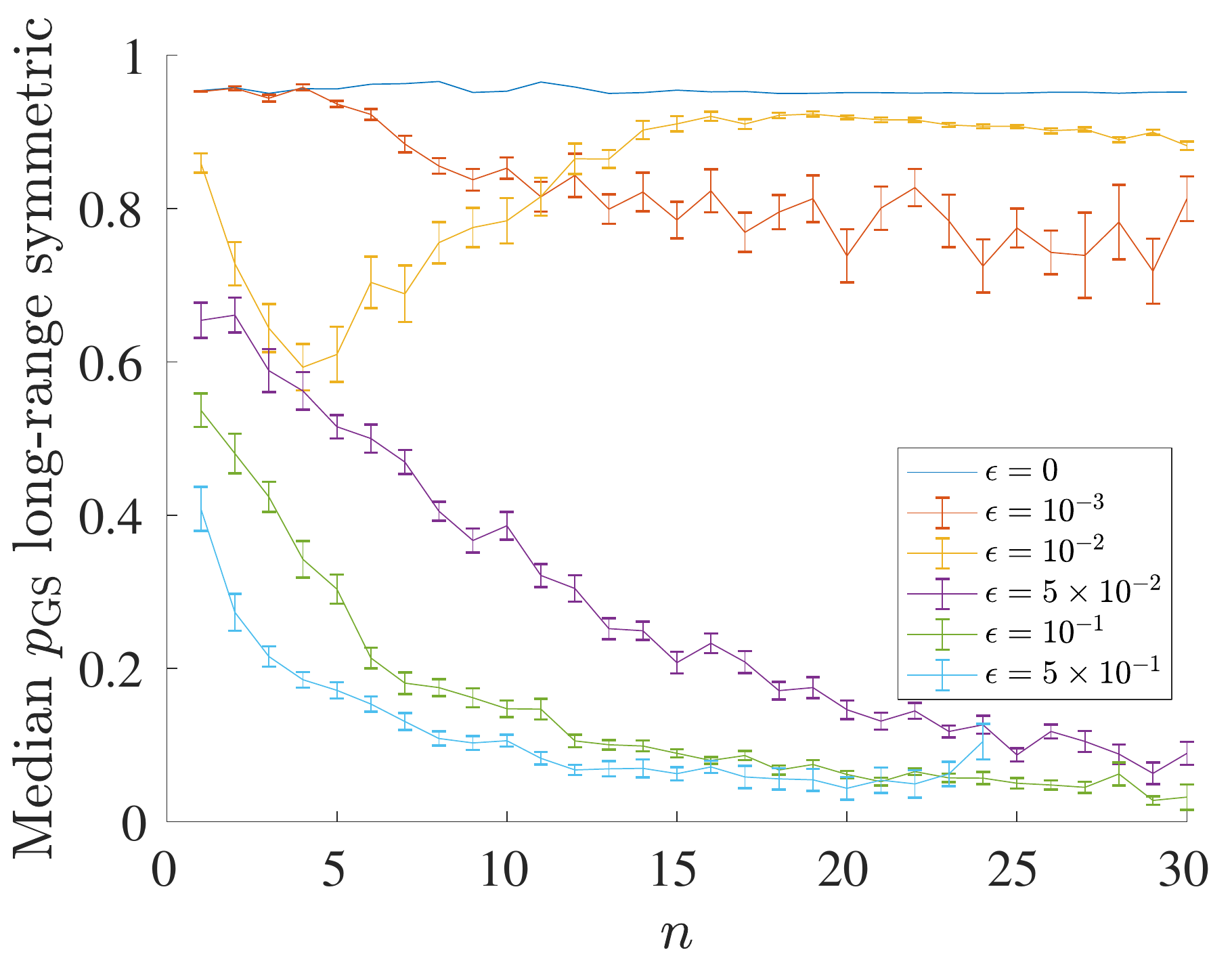}\label{fig:LSmanyeps}}
\subfigure[]{\includegraphics[width = \columnwidth]{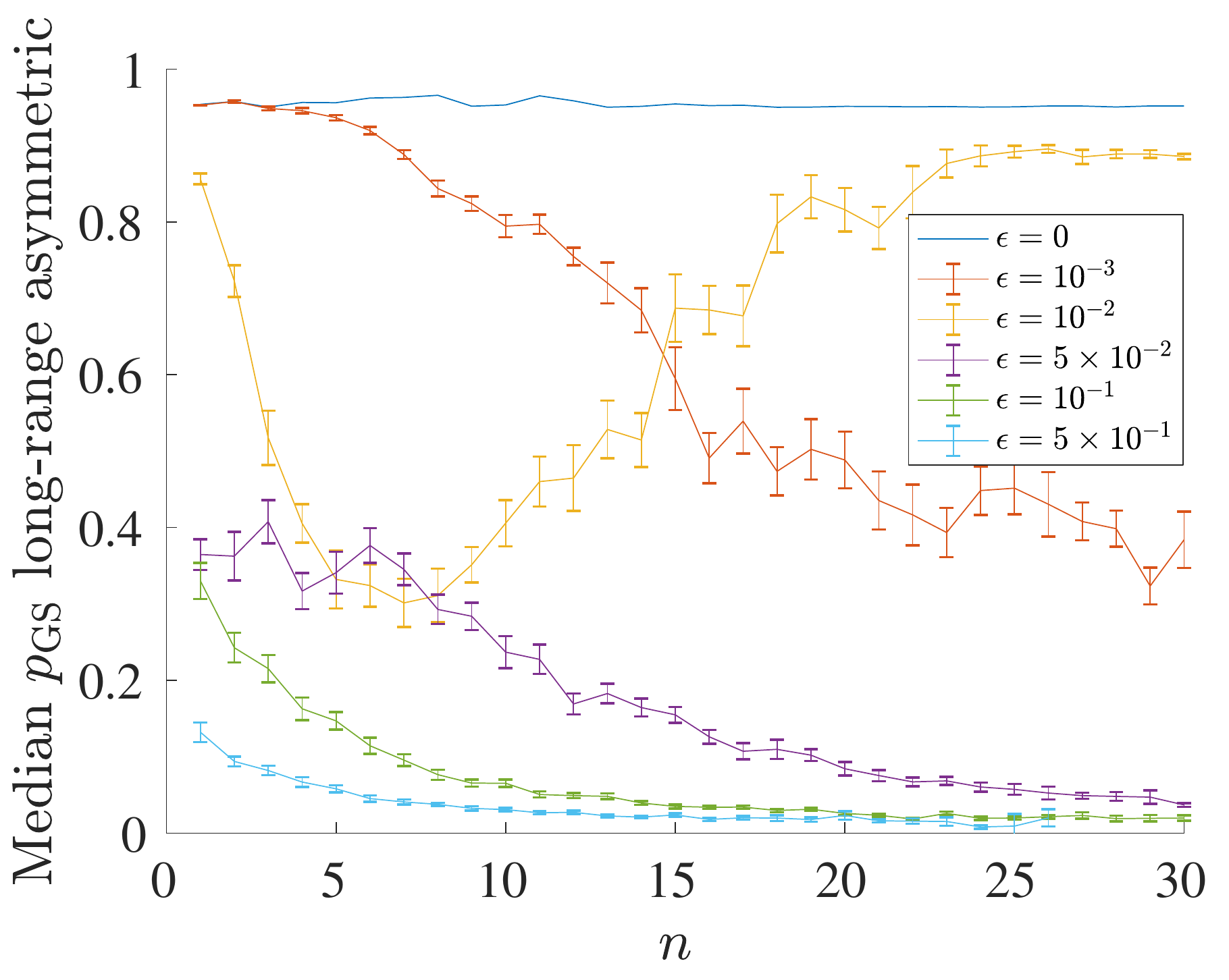}\label{fig:LAmanyeps}}
\subfigure[]{\includegraphics[width = \columnwidth]{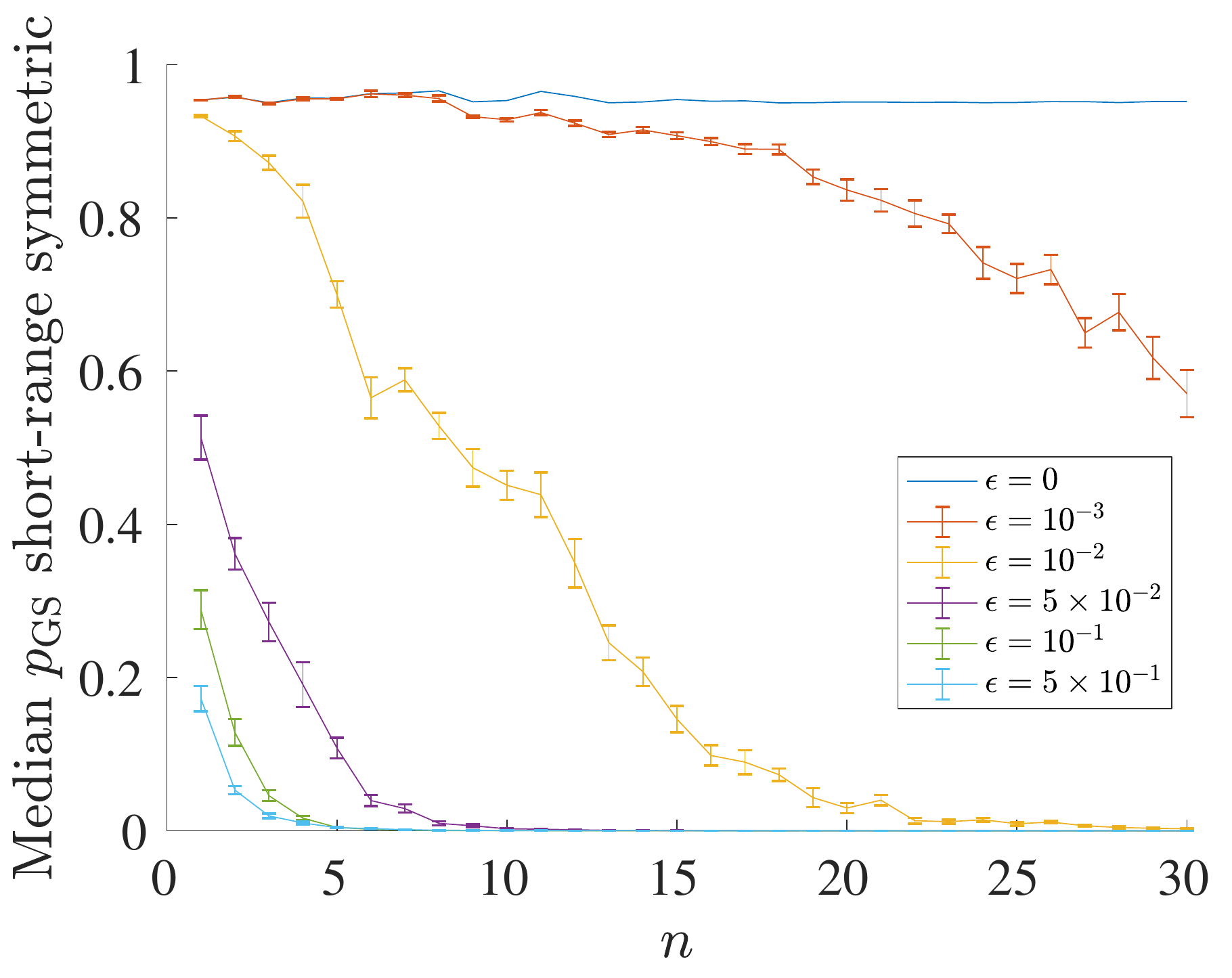} \label{fig:SSmanyeps}}
\subfigure[]{\includegraphics[width = \columnwidth]{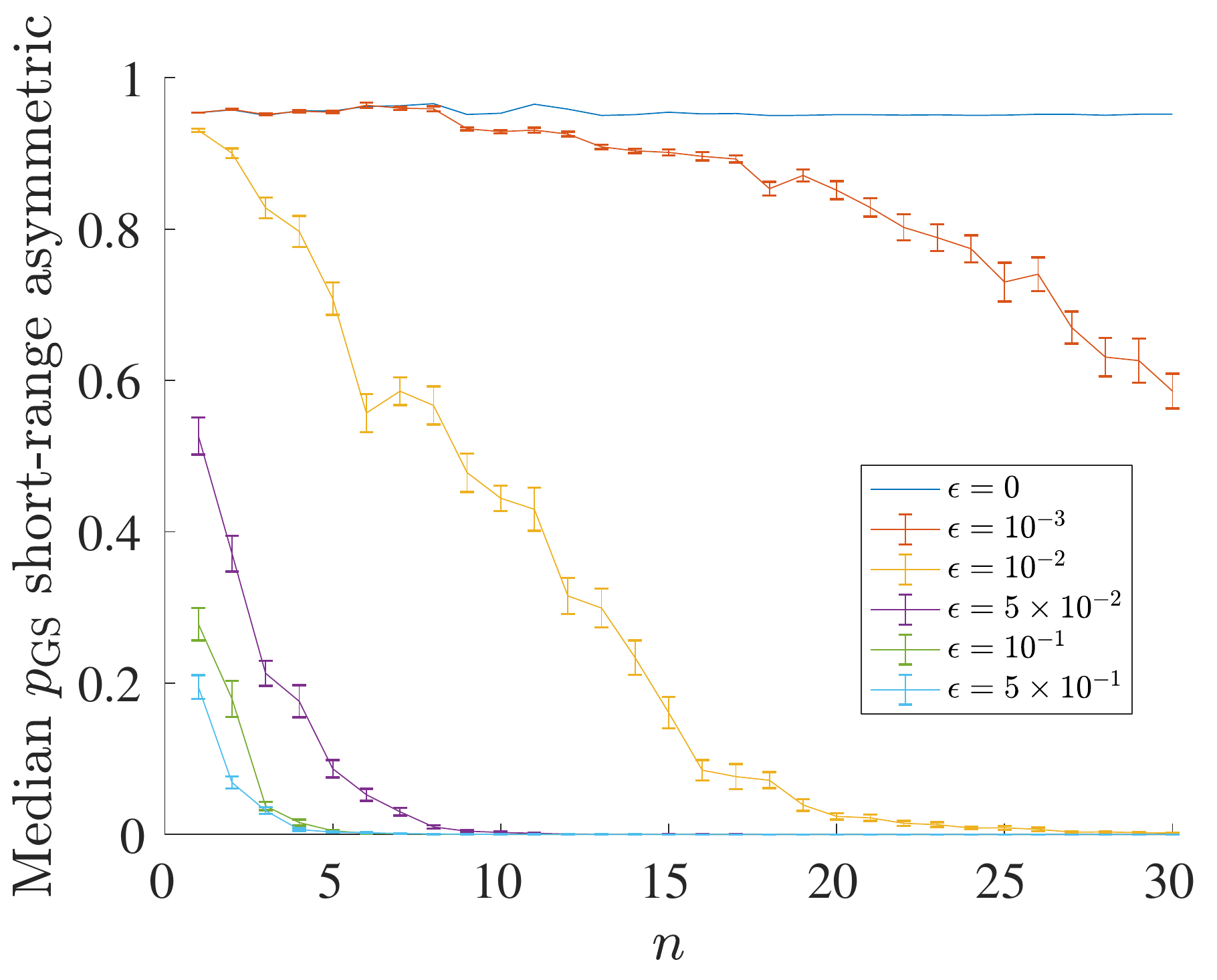}\label{fig:SAmanyeps}}
\caption{(Color online) The median success probability, $p_\mathrm{GS}$, at the end of an evolution of duration $t_f^\mathrm{Th}(n)$ as a function of $n$ for $\epsilon = 0, 10^{-3}, 10^{-2},  5\times 10^{-2}, 10^{-1}, 5\times 10^{-1}$. $\epsilon=0$ is the noiseless evolution ($\epsilon$ increases from top to bottom at $n=1$). $t_f^\mathrm{Th}(n)$ is chosen so that the success probability for the noiseless probability is just above $0.95$. The error bars are obtained by bootstrap sampling over $300$ realizations of the noise. (a) The long-range symmetric noise model. (b) The long-range asymmetric noise model. (c) The short-range symmetric noise model. (d) The short-range asymmetric noise model.}
\label{fig:mdnpgsvsnmanyeps}
\end{figure*} 

\section{Noise}\label{sec:noisemodels}

The noise models we consider here are phenomenological. They ignore the details of how the noise may be realized and instead posit some general properties that noisy systems might have. This method is especially well-suited to oracle algorithms. This is because oracles are typically very difficult to realize physically. Indeed, for the glued-trees problem we can show that the terms $H_0, H_1$, and $A$ in Eq.~\eqref{eq:QAHam} all need to be highly nonlocal and require experimentally difficult-to-engineer interactions (see Appendix~\ref{app:qubitgt}). Such oracle-level noise models are studied, e.g., in Refs.~\cite{Shenvi:03,temme2014runtime,cross2015quantum}, for circuit algorithms, and for some quantum walk algorithms (see~\cite{kendon2007decoherence} for a review), including the quantum walk version of the glued-trees algorithm~\cite{lockhart2014glued}. 

Our noise model is inspired by one due to Roland and Cerf~\cite{PhysRevA.71.032330}. The noise model they consider is a time-dependent random-matrix added to the Hamiltonian, with the entries of the random matrix being time-dependent random variables distributed as white noise with a cut-off. They show that this noise does not significantly affect the performance of the adiabatic algorithm as long as the cut-off frequency of the white noise is either much smaller or much greater than the energy scale of the noiseless Hamiltonian. They further explore this noise model in detail for the adiabatic Grover algorithm~\cite{Roland:2002ul}.

The noise models we study are as follows. We add a time-independent random matrix $h$ to our Hamiltonian $H(s)$ [Eq.~\eqref{eq:QAHam}]. We write our noisy Hamiltonian $\tilde{H}(s)$  as
\beq \label{eqt:QAErrorHam}
\tilde{H}(s) = H(s) + \epsilon h \ .
\eeq
We restrict the noise matrix $h$ to be inside the subspace spanned by the column basis, i.e., $h$ has the same dimensions as $H(s)$ when written in the column basis. The four noise models we consider correspond to different ways of choosing the random matrix $h$.

\subsection{Four noise models}
\label{sec:noisemodels}

We construct four noise models by selecting one branch in each of two dichotomies. The first dichotomy is between long-range and short-range noise models. The second dichotomy is between reflection-symmetric and reflection-asymmetric noise models. 

First we consider a noise model in which $h$ is chosen from the Gaussian Orthogonal Ensemble (GOE). This means that in any given basis, and in particular the column basis, the matrix elements of $h$ are distributed as
\beq \label{eq:goedef}
h_{ij} = h_{ji} =  \begin{cases} \mathcal{N}(0,1), \quad i \neq j \\
				 \mathcal{N}(0,2), \quad i = j
	    \end{cases} \ .
\eeq 
(Here Gaussian random variables are denoted as $\mathcal{N}(\mu,\sigma^2)$, with $\mu$ being the mean of the Gaussian and $\sigma$ the standard deviation.)
A standard way of generating such a matrix is to start with a matrix $M$ whose entries are independent $\mathcal{N}(0,1)$ random variables (and therefore non-Hermitian) and setting
\beq \label{eq:goegen}
h = \frac{M + M^T}{\sqrt{2}}.
\eeq
That Eq.~\eqref{eq:goedef} is obtained from Eq.~\eqref{eq:goegen} can be seen from the fact that the addition of two independent Gaussian random variables obeys 
\beq
\label{eq:Gauss-sum-diff}
\mathcal{N}(\mu_1,\sigma_1^2) \pm \mathcal{N}(\mu_2,\sigma_2^2) = \mathcal{N}(\mu_1 \pm \mu_2, \sigma_1^2 + \sigma_2^2), 
\eeq
combined with $a \mathcal{N}(\mu,\sigma^2) = \mathcal{N}\left(\mu,(a\sigma)^2\right)$ (see Appendix~\ref{app:GOE} for more details about the GOE).

We call this noise model---i.e., the model generated by adding a time-independent random matrix chosen from the GOE---the \emph{long-range asymmetric} (LA) noise model and denote $h$ in this case by $h_\mathrm{LA}$. \emph{Long-range} refers to the fact that $h$ contains matrix elements which connect all columns to all columns, and \emph{asymmetric} refers to $h$ breaking the reflection symmetry of the Hamiltonian. To see that $h$ breaks the reflection symmetry, notice that $h$ is not invariant under conjugation with the permutation matrix $P$, which, together with the fact that $h$ is time-independent, yields that $\tilde{H}$ is not reflection symmetric.

Next, we consider what we call the \emph{long-range symmetric} (LS) noise model. This noise model preserves the reflection symmetry of the problem. To generate this noise model, we first pick a matrix $\omega$ from the GOE. Then we reflection-symmetrize it:
\beq
h_\mathrm{LS} \equiv \frac{\omega + P \omega P}{\sqrt{2}}.
\eeq
Now $h_\mathrm{LS}$ is manifestly reflection symmetric and therefore so is $\tilde{H} = H + \epsilon h_\mathrm{LS}$. Note that $h_\mathrm{LS}$ still contains terms connecting distant columns.
From the above definition, we can check that the matrix elements of $h_\mathrm{LS}$ are distributed as
\begin{align} \label{eq:hLSdef}
&(h_\mathrm{LS})_{ij} = (h_\mathrm{LS})_{ji} \\ \nonumber
&= (h_\mathrm{LS})_{(2n+1)-i,(2n+1)-j} = (h_\mathrm{LS})_{(2n+1)-j,(2n+1)-i} \\ \nonumber
&= \begin{cases} \mathcal{N}(0,1), \quad i \neq j \\
				 \mathcal{N}(0,2), \quad i = j.
	    \end{cases}
\end{align} 
The reflection symmetry of $h_\mathrm{LS}$ implies that the spectrum of $\tilde{H}$ is reflection symmetric as well in this case. 

We next turn to the short-range noise models. These noise-models only connect neighboring columns in the glued-trees graph. We examine both \emph{short-range asymmetric} (SA) and \emph{short-range symmetric} (SS) noise models. In the SA model, the Hamiltonian perturbation is given by
\beq
(h_\mathrm{SA})_{ij} = \begin{cases} \mathcal{N}(0,1) \quad \abs{i-j}=1, \\
									0	   \quad \mathrm{otherwise}.
				   \end{cases}
\eeq
In the SS model, we have
\beq
h_\mathrm{SS} = \frac{h_\mathrm{SA} + Ph_\mathrm{SA}P}{\sqrt{2}},
\eeq
which preserves the reflection symmetry of the Hamiltonian.

We remark that the parameter controlling the strength of the noise $\epsilon$, can be absorbed into the standard deviations of the Gaussian random variables: e.g., $\epsilon \mathcal{N}(0,1) = \mathcal{N}(0,\epsilon^2)$. Therefore, the larger the noise, the greater the spread of the Gaussians according to which the matrix elements are drawn. 

\begin{figure}[t]
\centering
\includegraphics[width=\columnwidth]{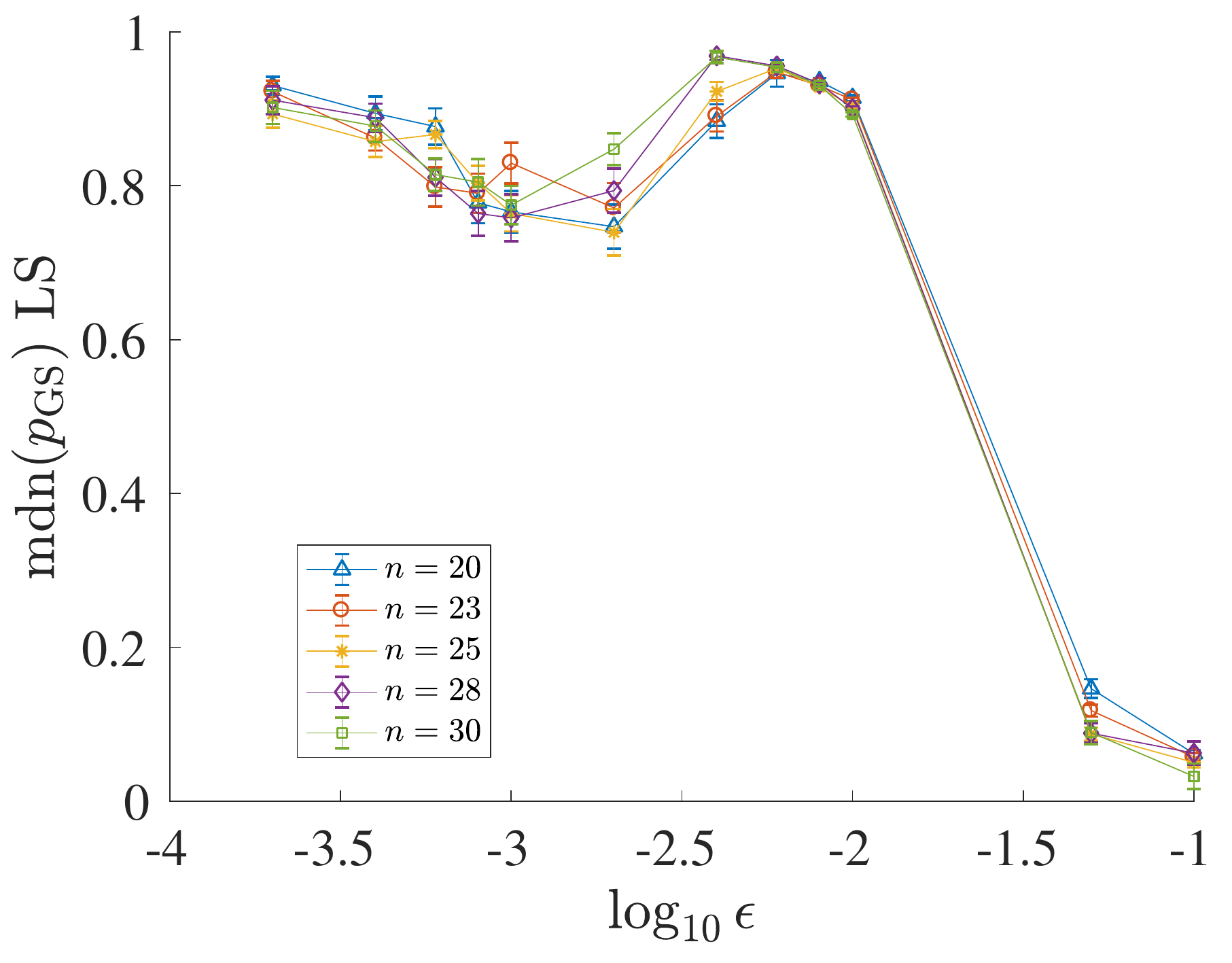}
\caption{(Color online) Median success probability vs. the log (base 10) of the strength of the noise for the LS model for several, larger values of the problem size $n$. There is a fall, then a rise, and then again a fall in this success probability as a function of $\epsilon$ for all  values of $n$ displayed.}
\label{fig:mdnpgsvsepsmanyn}
\end{figure}

\begin{figure*}[t]
\subfigure[]{\includegraphics[width = \columnwidth]{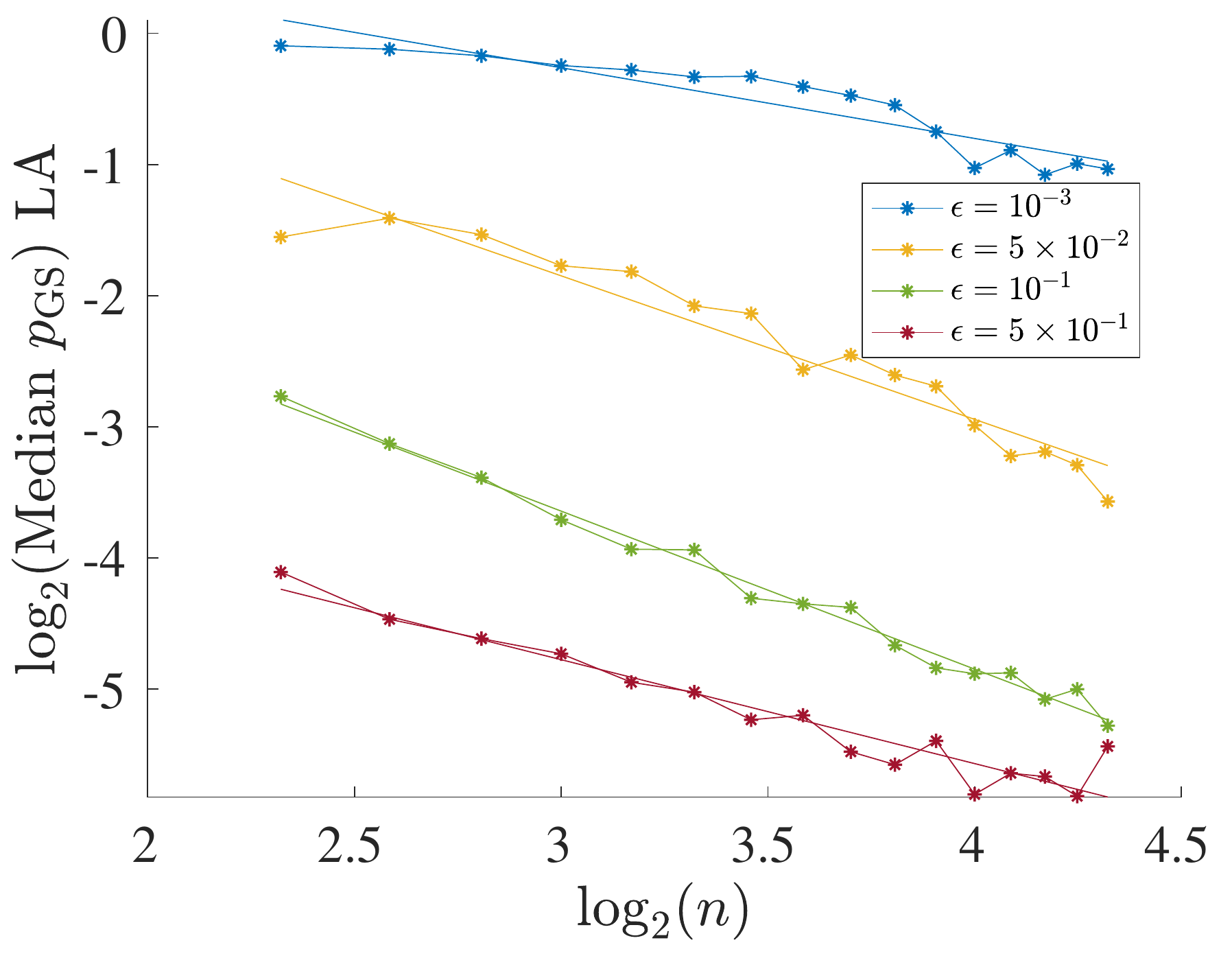}\label{fig:polyfitLSmanyeps}}
\subfigure[]{\includegraphics[width = \columnwidth]{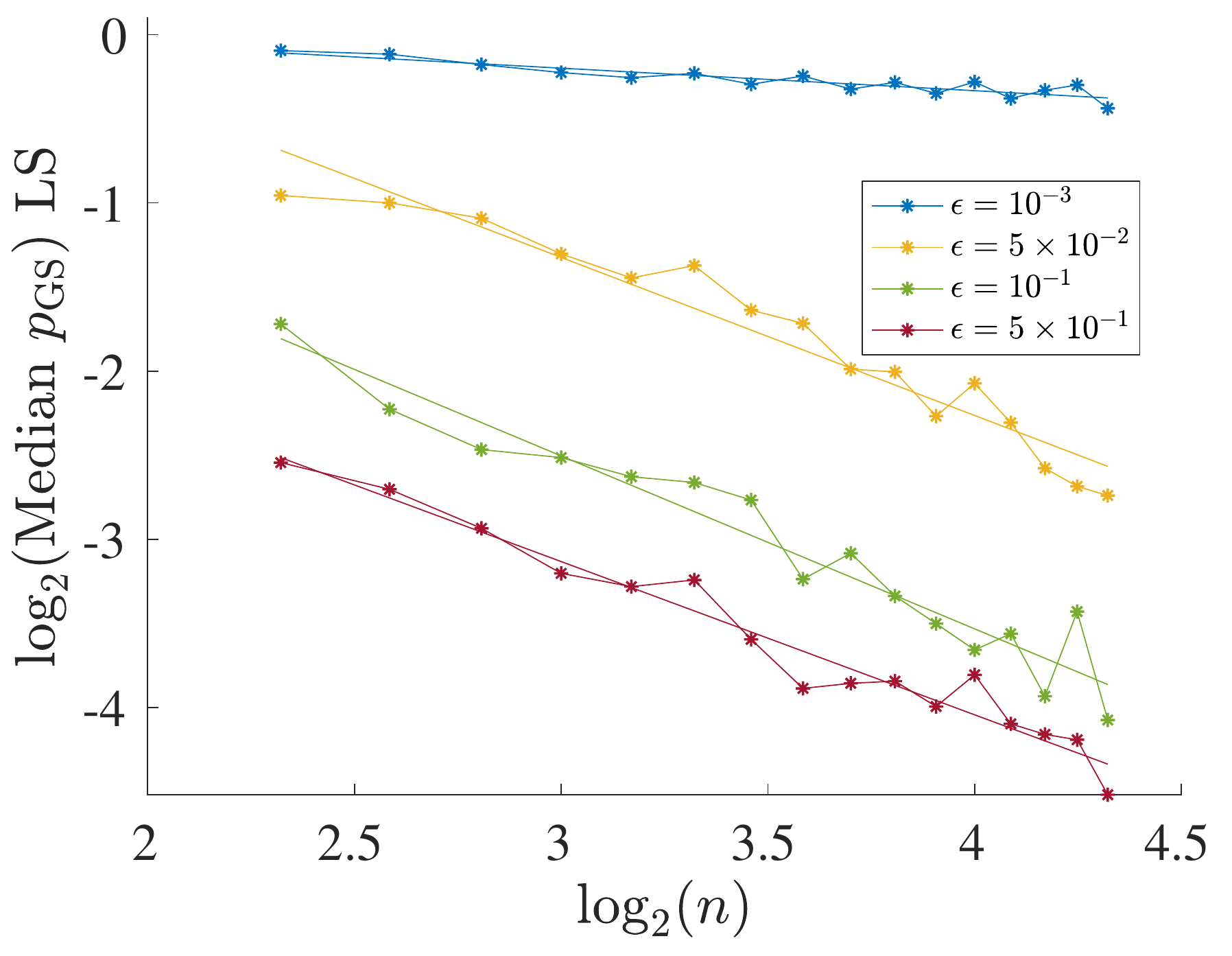}\label{fig:polyfitLAmanyeps}}
\subfigure[]{\includegraphics[width = \columnwidth]{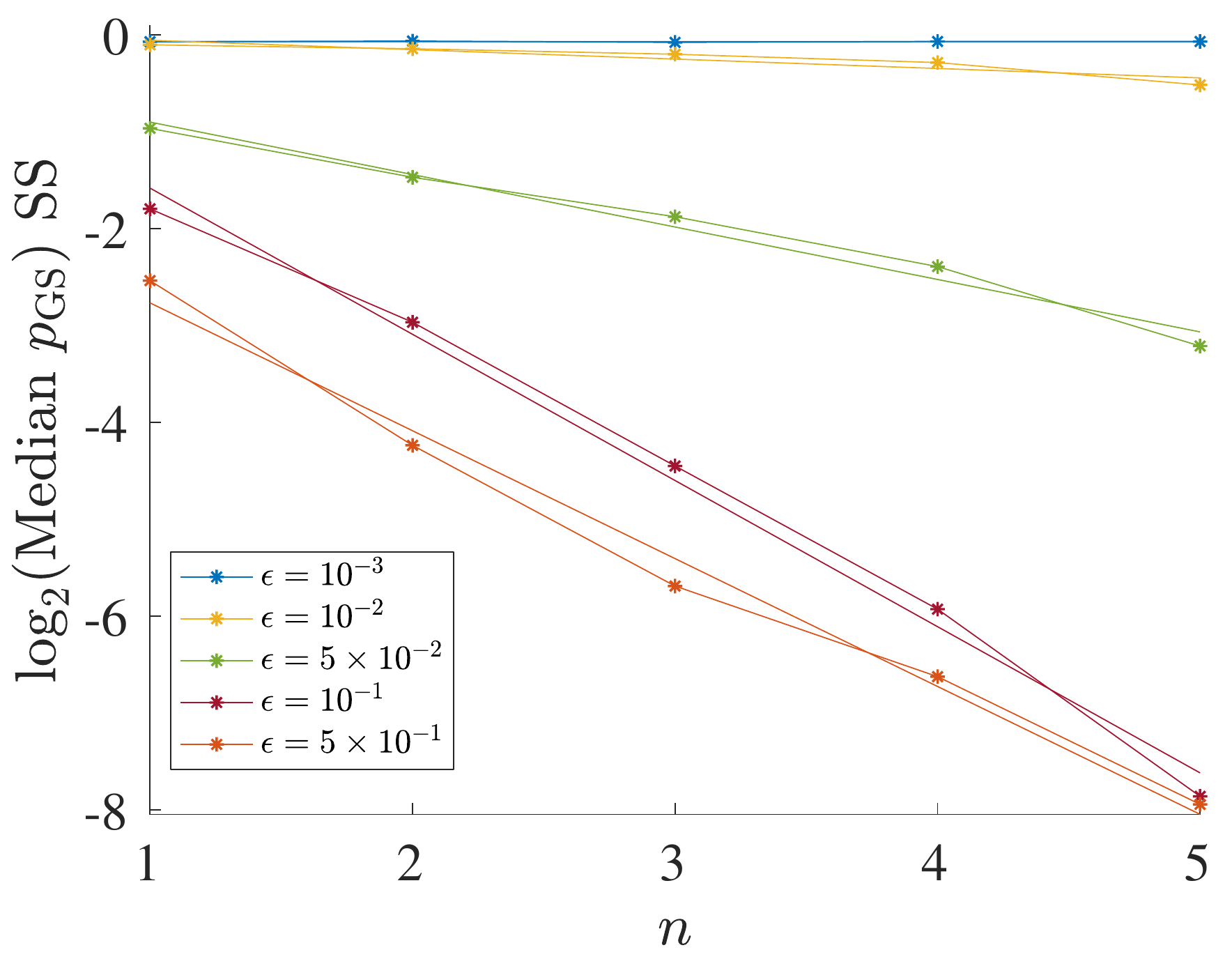} \label{fig:expfitSSmanyeps}}
\subfigure[]{\includegraphics[width = \columnwidth]{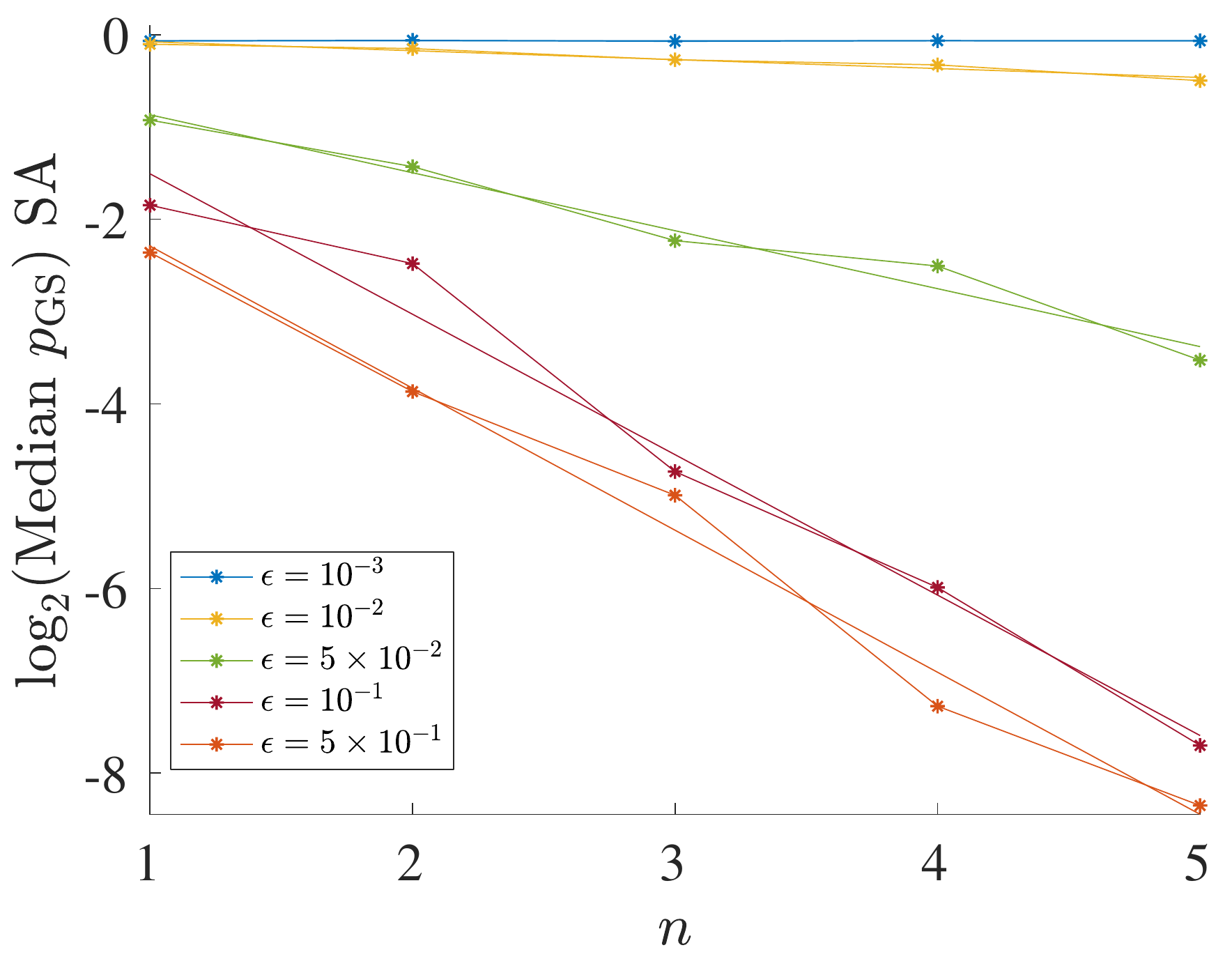}\label{fig:expfitSAmanyeps}}
\caption{(Color online) Base-2 logarithm of the median ground state success probability, as a function of problem size $n$, or the logarithm of the problem size $\log_2 n$, at different noise levels $\epsilon$. Polynomial fits of the form $\mathcal{O}(n^\alpha)$ are performed for the long-range models. Exponential fits of the form $\mathcal{O}(2^{\alpha n})$ are performed for the short range models. This is done for $\epsilon \in \{10^{-3},10^{-2}, 5\times 10^{-2}, 10^{-1}, 5\times 10^{-1}\}$ for the short-range models ($\epsilon$ increases from top to bottom at $n=1$). For the long-range models, the set of values of $\epsilon$ is the same as for the short-range models, except that we omit the $\epsilon = 10^{-2}$ case since it shows anomalous behavior (i.e., a rise and a fall) which doesn't fit a decay. The scaling coefficient $\alpha$ as a function of the noise $\epsilon$ is shown in Fig.~\ref{fig:scalingcoeffsvseps}. (a) The long-range symmetric noise model. (b) The long-range asymmetric noise model. (c) The short-range symmetric noise model. (d) The short-range asymmetric noise model.}
\label{fig:expfitmdnpgsvsnsymmanyeps}
\end{figure*} 

\begin{figure}[t]
\centering
\includegraphics[width=\columnwidth]{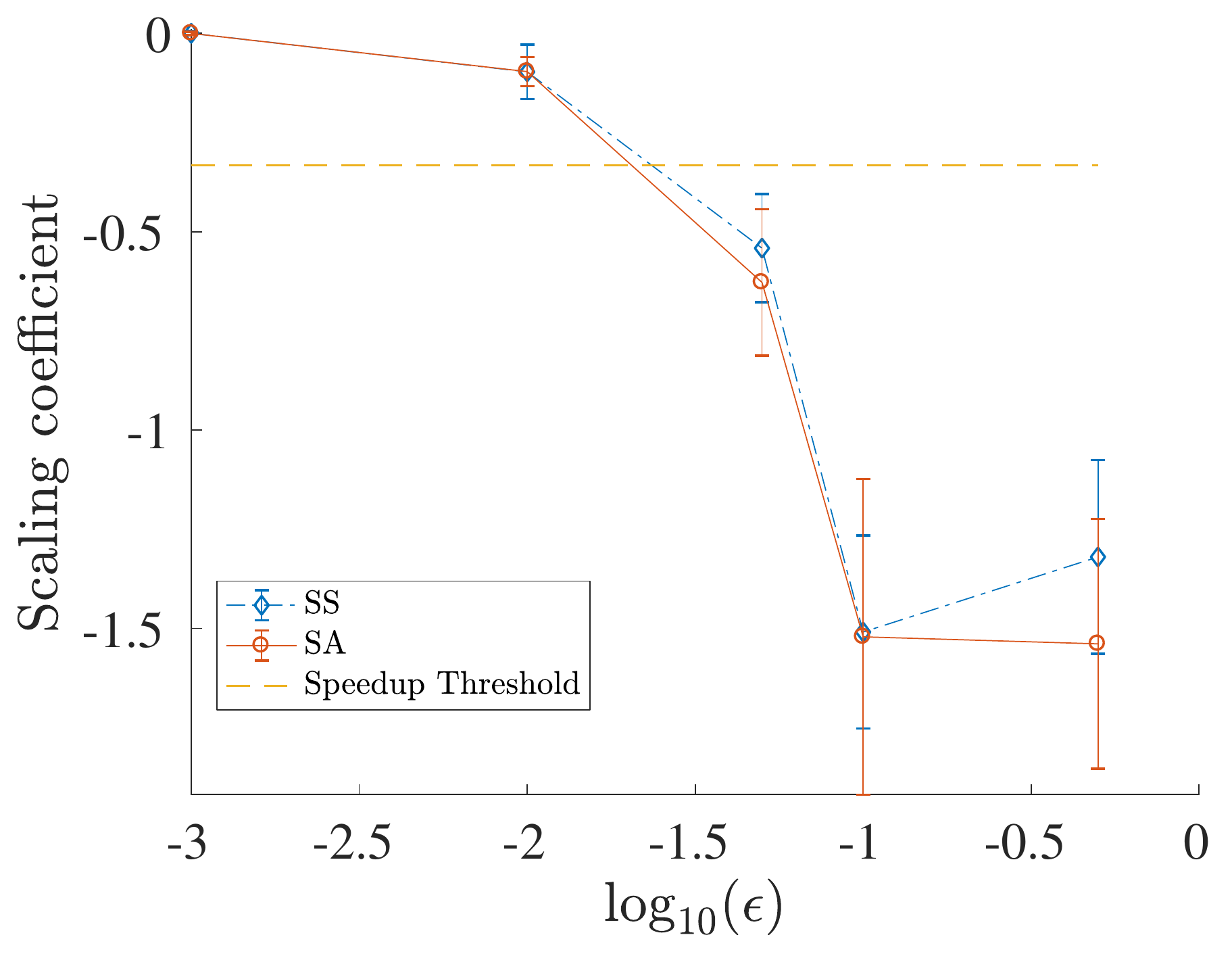}
\caption{(Color online) The exponential scaling coefficient $\alpha$ as a function of the base-10 logarithm of the strength of the noise $\epsilon$ for the short-range noise models. (The long-range noise models do not show an exponential scaling.) The scaling coefficient $\alpha$ is obtained by performing an exponential fit of the form $\mathcal{O}(2^{\alpha n})$, to the median success probability, $p_\mathrm{GS}(t_f^\mathrm{Th})$ vs. $n$ curves [shown in Figs.~\ref{fig:expfitSSmanyeps} and~\ref{fig:expfitSAmanyeps}]. The dashed horizontal line at $\alpha = -1/3$ represents the scaling coefficient below which the speedup, over the best possible classical algorithm to solve the glued-trees problem, is lost. Thus, the short-range models lose the speedup for $\epsilon \gtrsim 10^{-1.75}$.  The error-bars are 95\% confidence intervals. While some error-bars are large due to the fits being performed on a small number of data points, the trend in the data is clear.}
\label{fig:scalingcoeffsvseps}
\end{figure}

\begin{figure*}[t]
\subfigure[]{\includegraphics[width = \columnwidth]{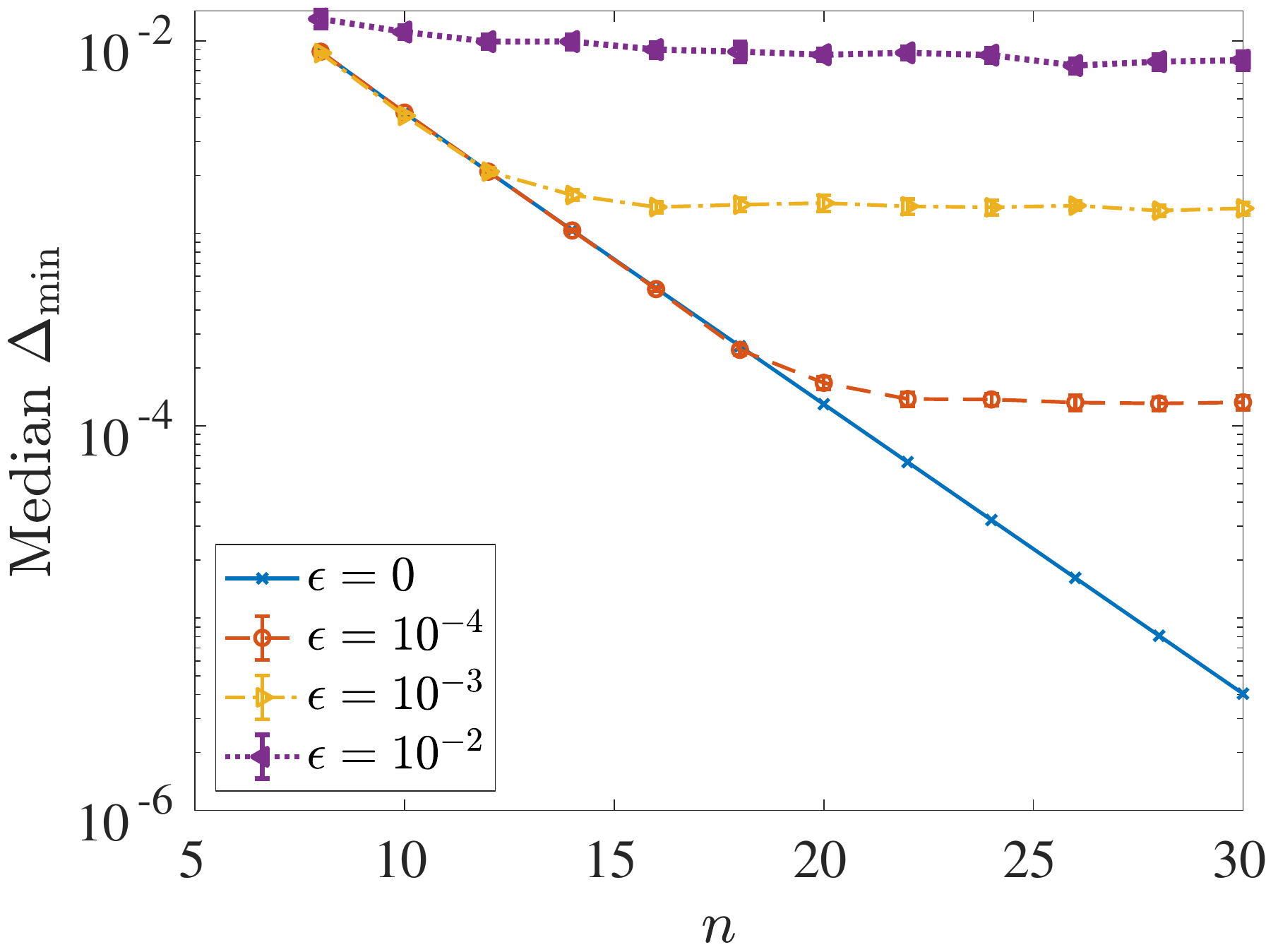}\label{fig:LSgaps}}
\subfigure[]{\includegraphics[width = \columnwidth]{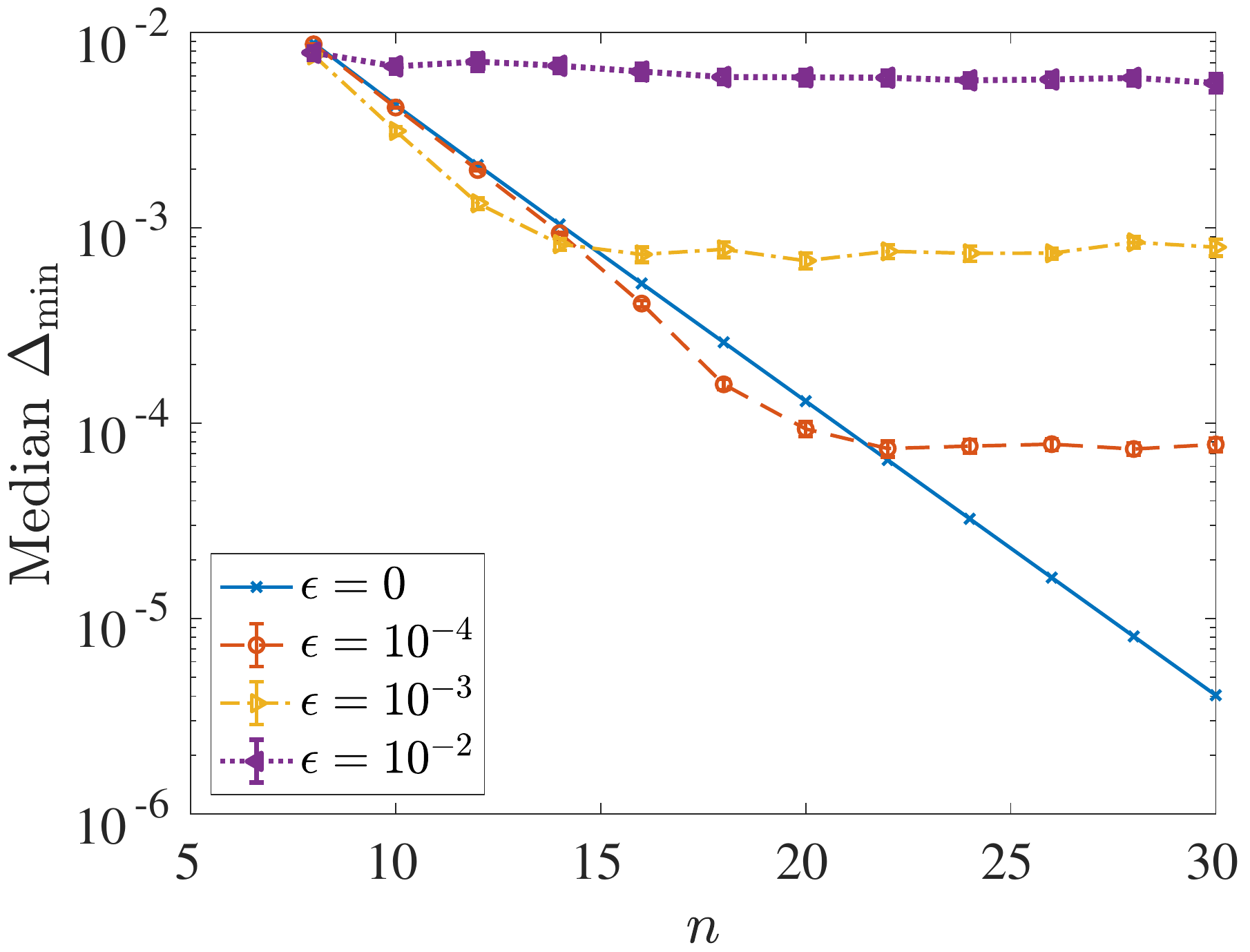}\label{fig:LAgaps}}
\subfigure[]{\includegraphics[width = \columnwidth]{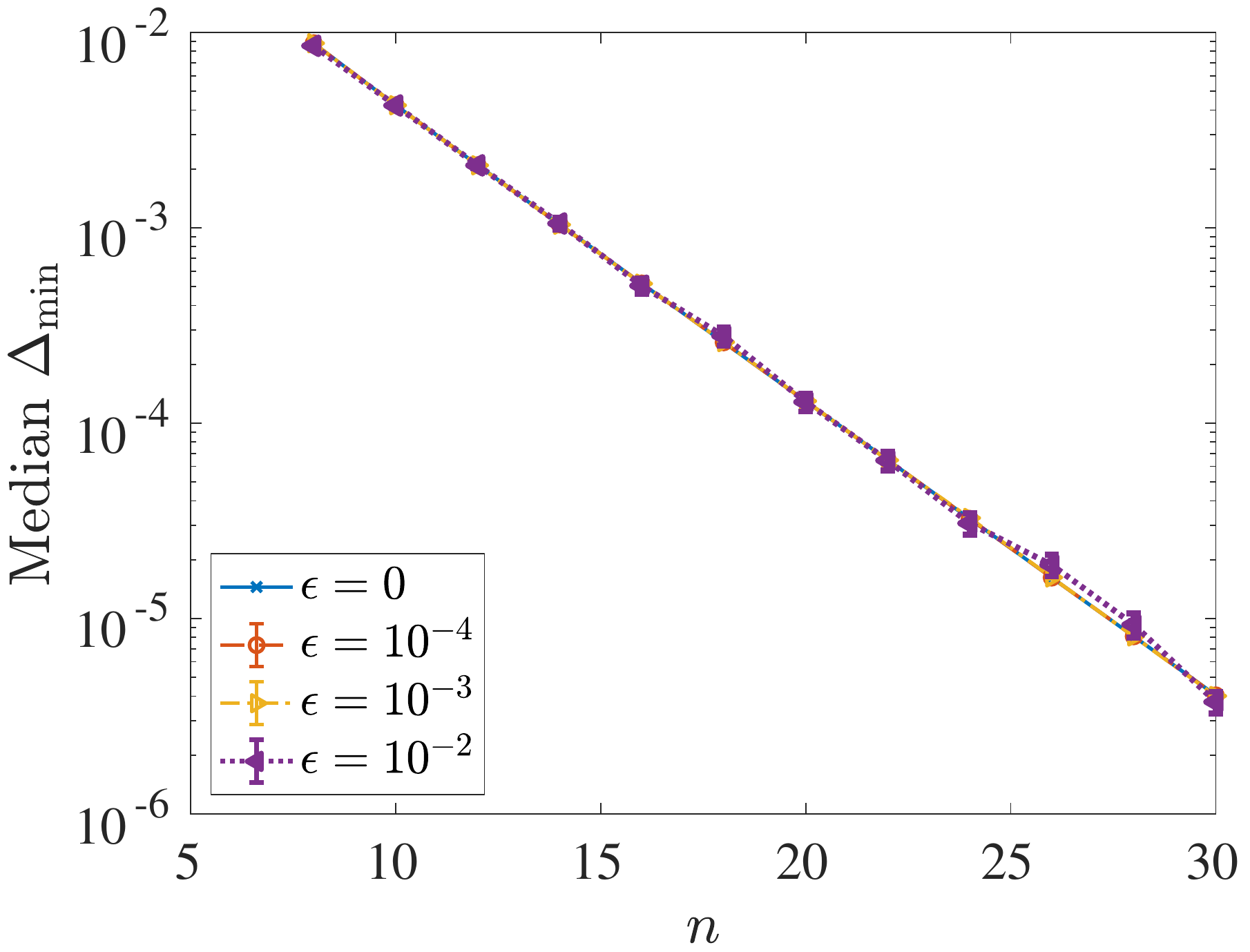} \label{fig:SSgaps}}
\subfigure[]{\includegraphics[width = \columnwidth]{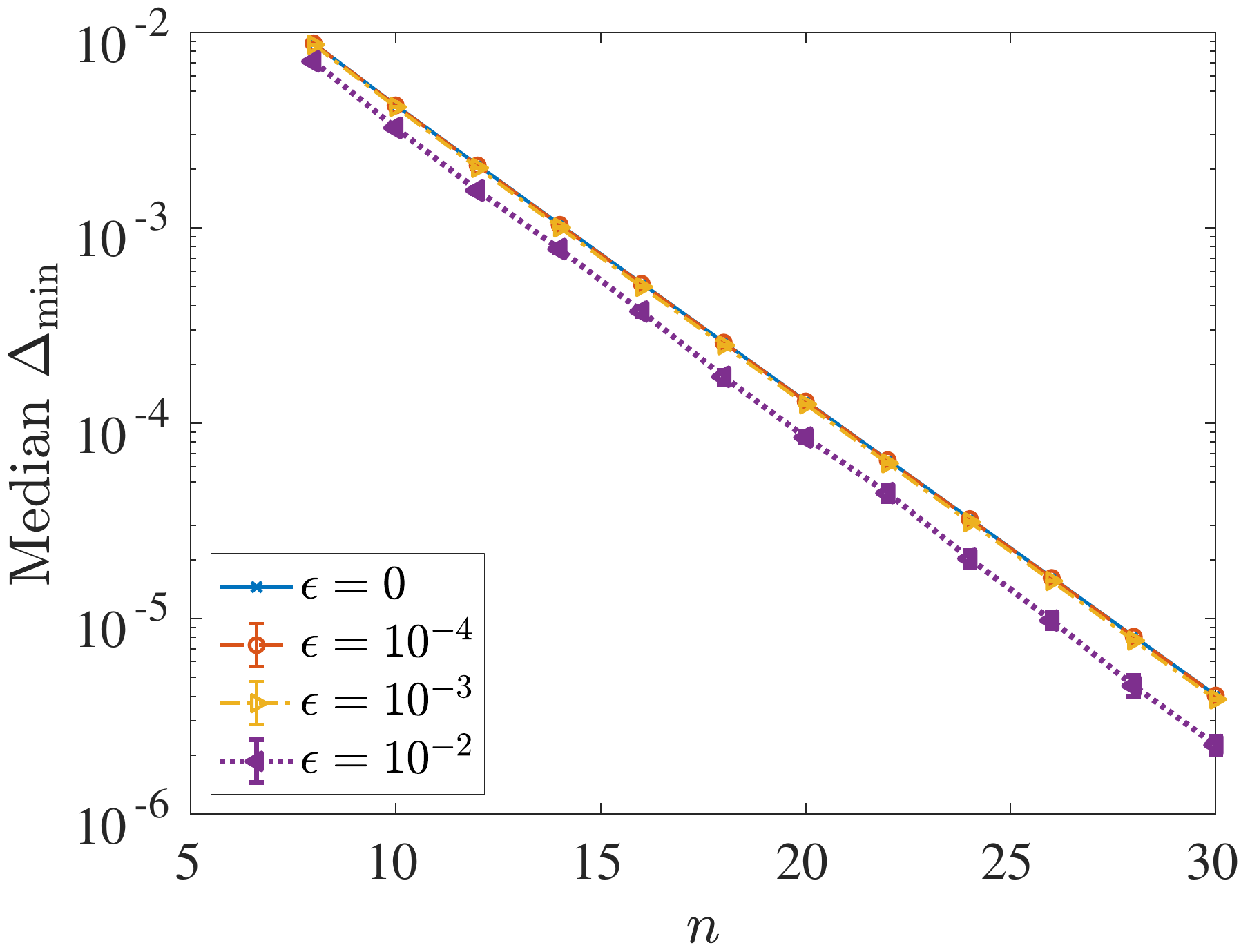}\label{fig:SAgaps}}
\caption{(Color online) The scaling of the median minimum gap (log-scale) as a function of problem size $n$ for the four noise models with $\epsilon \in \{10^{-4},10^{-3},10^{-2}\}$. The solid blue line represents the noiseless algorithm ($\epsilon = 0$), which has an exponentially closing minimum gap. (a) Long-range symmetric. (b) Long-range asymmetric. (c) Short-range symmetric. (d) Short-range asymmetric. The long-range models display a constant gap with problem size, while the short-range models exhibit an exponentially decreasing gap with problem size. This is consistent with the perturbative argument given in Eq.~\eqref{eq:gap-LA}.}
\label{fig:noisygapsvsn}
\end{figure*} 

\section{Noisy glued-trees: Results from numerical simulations}\label{sec:numericalresults}

We simulate the Schr{\"o}dinger evolution
\beq
i \hbar \frac{d}{dt} \ket{\tilde{\psi}_{t_f}(s)} = t_f \tilde{H}(s) \ket{\tilde{\psi}_{t_f}(s)}
\eeq
using the different noise models and calculate the probability of finding the EXIT vertex at the end of the anneal, i.e., the success probability:
\beq
p_\mathrm{GS}[t_f^\mathrm{Th}(n)] \equiv \abs{\braket{\phi_0(s) | \tilde{\psi}_{t_f^\mathrm{Th}(n)}(1)}}^2 .
\eeq
We choose the annealing timescale $t_f$ to be equal to threshold timescale of the noiseless algorithm, i.e., the timescales depicted in Fig.~\ref{fig:gtnoiselessqa}. We do this because it is natural to imagine that one operates the algorithm in the regime in which the noiseless algorithm succeeds. Note that this probability is a random variable whose value depends on the specific noise realization, and we focus on the typical (median) value of this random variable. The median $p_\mathrm{GS}$ will depend on (a) the noise model, (b) the strength of the noise $\epsilon$, and (c) the problem size $n$.

In Fig.~\ref{fig:mdnpgsvsnmanyeps}, we plot the median success probability for the four different noise models specified in Sec.~\ref{sec:noisemodels} as a function of problem size $n$, equal to the depth of one of the two binary trees.

The first observation is that the median $p_\mathrm{GS}$ behavior is not significantly different between the symmetric and asymmetric noise model, for both the long-range and short-range variants, although the symmetric noise does slightly outperform the asymmetric case.  This suggests that while the symmetry of the spectrum is an important aspect of the performance of the noiseless algorithm (by allowing transitions from and back to the ground state), the noisy algorithm is somewhat robust to reflection symmetry-breaking.

Next, a remarkable feature of Fig.~\ref{fig:mdnpgsvsnmanyeps} is that for large enough $n$ ($n \gtrsim 13$), the success probability is non-monotonic with respect to the strength of the noise $\epsilon$. This can be seen more clearly in Fig.~\ref{fig:mdnpgsvsepsmanyn}: as expected, there is a fall in the success probability from $\epsilon=0$ to $\epsilon=10^{-3}$, but, surprisingly, there is a rise from $\epsilon=10^{-3}$ to $\epsilon=10^{-2}$. For higher values of $\epsilon$, the probability falls off, again as expected. We explain the counterintuitive rise in the next section.  

We also examine, for a given noise model and a given noise strength, whether the quantum speedup of the noiseless algorithm is retained. Since $2^{n/3}$ is the best possible time scaling that a noiseless classical random walk can achieve (see Sec.~\ref{sec:gtproblem} and Ref.~\cite{childs2003exponential}), 
the $p_\mathrm{GS}(t_f^\mathrm{Th})$ vs. $n$ scaling should decline no faster than $2^{-n/3}$ for a quantum speedup to persist.
But, it might not be an exponential speedup: if the success probability for the noisy quantum algorithm declines as an exponential function that decreases slower than $2^{-{n/3}}$, then the speedup over the classical algorithm will instead be a polynomial speedup.

We thus perform fits to the $p_\mathrm{GS}(t_f^\mathrm{Th})$ vs. $n$ curves, displayed in Fig.~\ref{fig:expfitmdnpgsvsnsymmanyeps}. For the long-range models, we exclude small values of $n$ because the asymptotic behavior starts to show only at larger values of $n$. For the short-range models, we only perform the fits on small values of $n$ because exponential decay causes the values of the success probability to be very small for larger values of $n$. The long-range models fit polynomials of the form $\mathcal{O}(n^\alpha)$. This means that the long-range models have an exponential speedup over the noiseless classical algorithm. (We argue below why in fact referring to this as a quantum speedup is misleading.) On the other hand, the short-range models fit exponentials of the form $\mathcal{O}(2^{\alpha n})$, which means that they do not exhibit a polynomial speedup over the noiseless classical algorithm if $\alpha < -1/3$ (as discussed in the previous paragraph). In Fig.~\ref{fig:scalingcoeffsvseps}, we plot the scaling coefficient $\alpha$ as a function of the noise strength $\epsilon$ for the short-range noise models. As is apparent from Fig.~\ref{fig:scalingcoeffsvseps}, the quantum speedup is lost for the short-range models for $\epsilon \gtrsim 10^{-1.75}\approx 0.018$. 

Let us now explain why the exponential speedup exhibited by the long-range noise models is misleading. For this speedup to count as a genuine quantum speedup, we must compare the quantum algorithm with an appropriate classical algorithm, so that we do not bias our analysis in favor of the quantum algorithm. How do we construct the appropriate classical algorithm in this case? Recall that in the quantum case, the long-range noise Hamiltonians have $\mathcal{N}(0,1)$-distributed off-diagonal terms. These terms connect distant columns of the graph. Thus, these models ought to be compared with classical random walks which have $\mathcal{N}(0,1)$-distributed transition probabilities between distant nodes of the graph. This needs to be normalized by a factor that is $\mathcal{O}(n)$ since there are $2n+2$ columns in total. Therefore, at any given time-step, in whichever column the random walker is, there is an $\mathcal{O}(1/n)$ probability that the random walker will transition directly to the EXIT vertex in the next time-step. Hence such a classical random walk will land at the EXIT vertex in $\mathcal{O}(n)$ time, and compared to the appropriate classical algorithm, the quantum algorithm with the long-range noise does not have an exponential speedup.

\begin{figure*}[t]
\centering
\subfigure[]{\includegraphics[width=0.66\columnwidth]{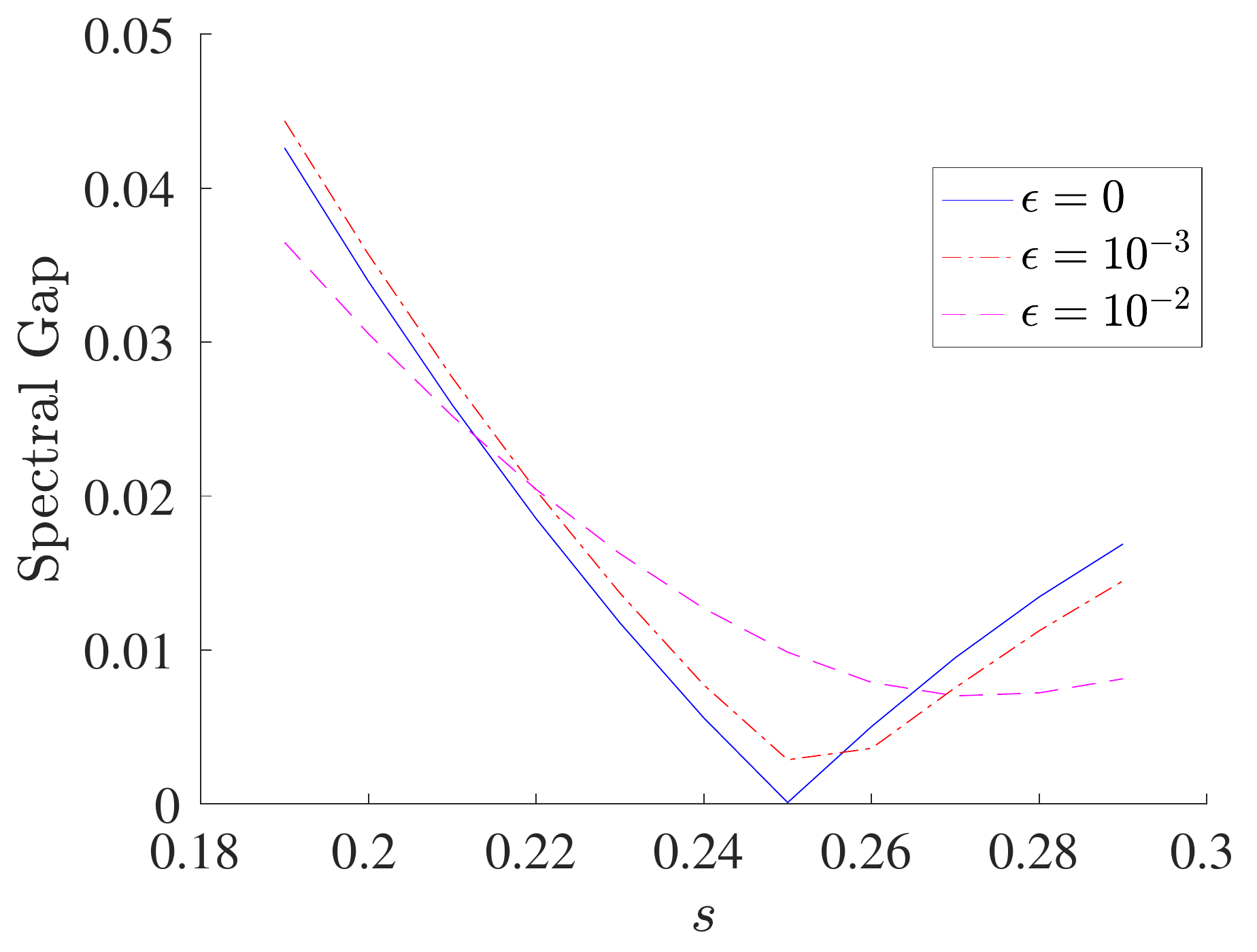}\label{fig:typicalspectrum}} 
\subfigure[]{\includegraphics[width=0.66\columnwidth]{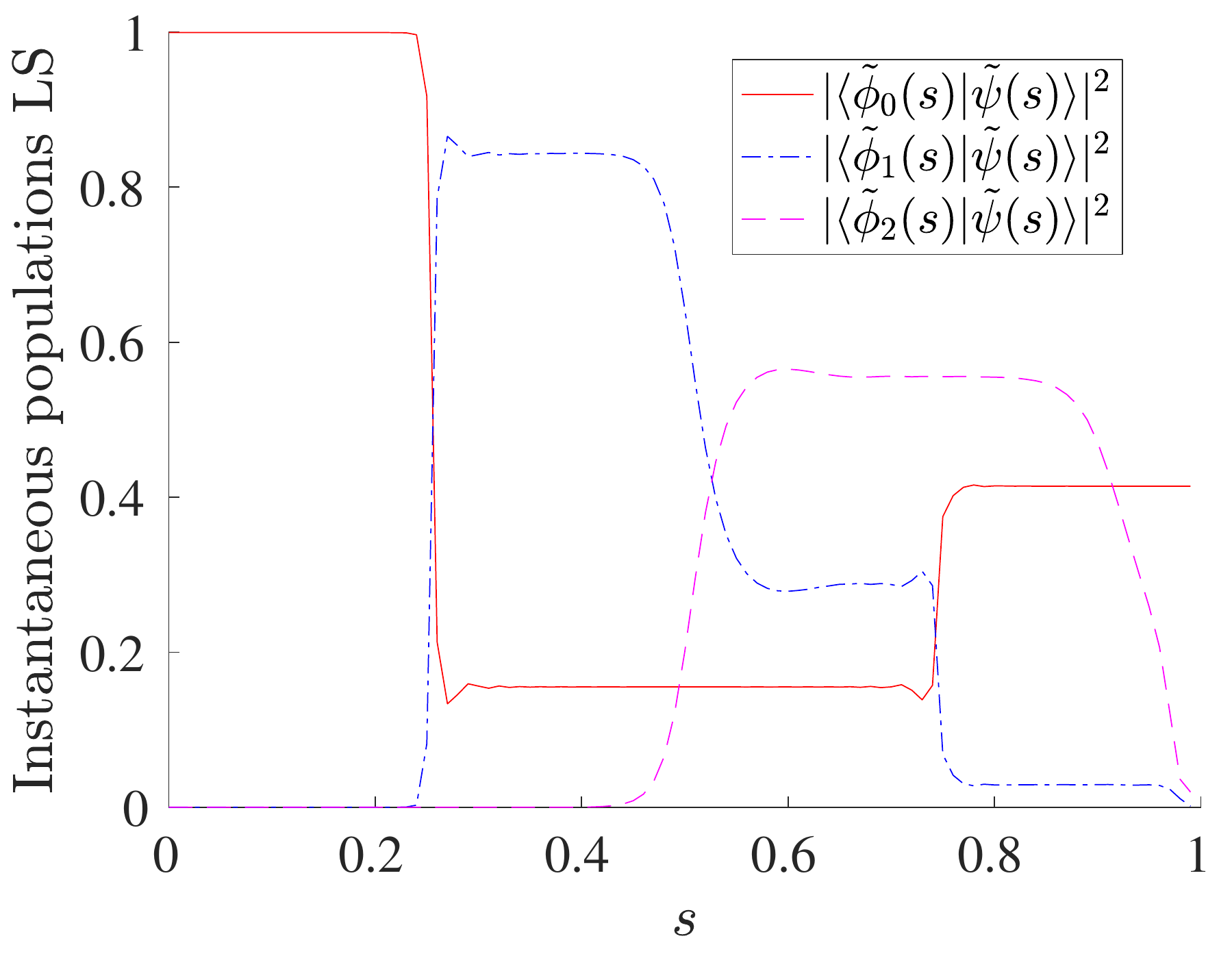}\label{fig:10-3dynamics}}
\subfigure[]{\includegraphics[width=0.66\columnwidth]{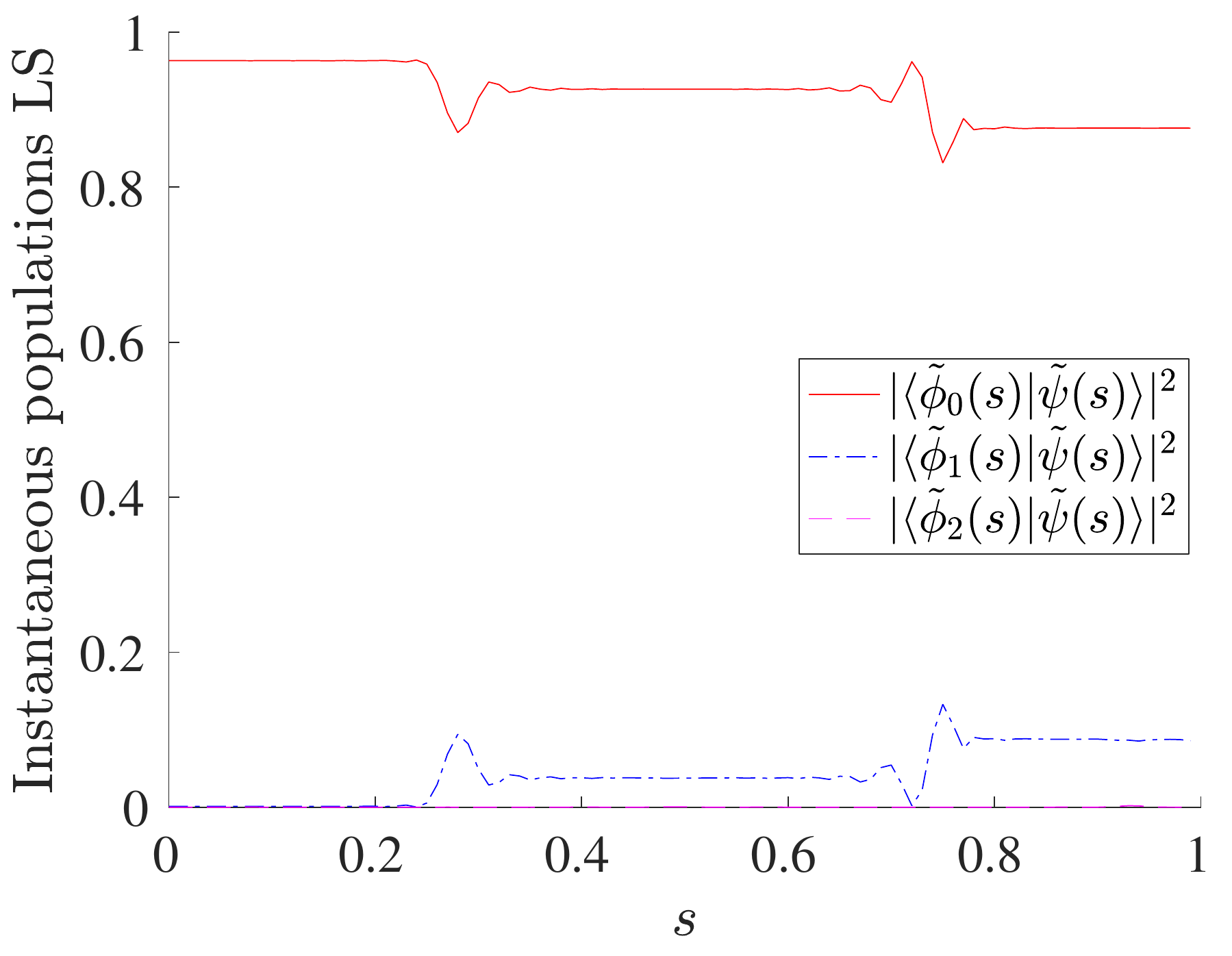}\label{fig:10-2dynamics}}
\caption{(Color online) Two random instances of the long-range symmetric noise for $n=20$ and $t_f = 12125$, with $\epsilon=10^{-3}$ and $\epsilon=10^{-2}$. (a) The spectral gap between the ground state and the first excited state for the two noisy instances and the noiseless case ($\epsilon = 0$), around $s = s^*$, i.e., near the location of the first exponentially closing gap in the noiseless problem. Both noisy instances increase the spectral gap from its value in the noiseless problem, with the increase being larger in the $\epsilon = 10^{-2}$ case. (b) The populations in the lowest three eigenstates of the noisy Hamiltonian as a function of the anneal parameter $s$, for $\epsilon = 10^{-3}$. The diabatic transitions do not happen as cleanly as in the noiseless case [shown in Fig.~\ref{fig:noiselesspopsn20}].  (c) As in (b), for $\epsilon=10^{-2}$. The dynamics are very close to adiabatic. }
\label{fig:popdynamicssymnoise}
\end{figure*}

\section{Noise-induced adiabaticity}\label{sec:explanation}

Two things stand out in the results presented in the previous section. The first is that for the long-range models, there is a rise in success probability from $\epsilon \approx 10^{-3}$ to $\epsilon \approx 10^{-2}$. The second is that the long-range models have an exponential speedup over the noiseless classical algorithm for a wide range of noise strengths, when compared to the short-range models. In this section, we will provide explanations for these two observations.

One phenomenon helps explain both these behaviors: In the long-range models, the noise typically leads to a larger spectral gap. To see this, consider what happens to the spectral gap at first-order in perturbation theory when we add long-range noise. Consider first the asymmetric case. Let $E_0^{(1)}(s)$ and $E_1^{(1)}(s)$ be the first-order corrections to the ground- and first excited state respectively. We know that
\bes
\begin{align}
E_0^{(1)}(s) &= \epsilon \braket{\phi_0 (s)| h_\mathrm{LA}(s) | \phi_0(s)} \\
&\sim \epsilon \mathcal{N}(0,2),
\end{align}
\ees
where $\ket{\phi_0(s)}$ represents the ground state of the unperturbed (noiseless) problem at $s$. We have used the fact that $h_\mathrm{LA}(s)$ is drawn from the GOE, and that the diagonal elements of a GOE matrix are distributed with variance $2$ [see Eq.~\eqref{eq:goedef}]. A similar calculation gives that $E_1^{(1)}(s) \sim \epsilon \mathcal{N}(0,2)$. Putting these together, we can approximate the spectral gap of the perturbed Hamiltonian using first-order perturbation theory as follows.
\bes
\label{eq:gap-LA}
\begin{align}
\tilde{\Delta}_\mathrm{LA}(s) &= \tilde{E}_1(s) - \tilde{E}_0(s)  \\
&\approx [ E_1(s) + \epsilon E_1^{(1)}(s) ] - [ E_0(s) + \epsilon E_0^{(1)} (s) ] \\
&= \Delta(s) + \epsilon \mathcal{N}(0,4), \label{eq:gappert}
\end{align}
\ees
where $\Delta(s)$ is the spectral gap of the noiseless problem. To obtain the last line we used Eq.~\eqref{eq:Gauss-sum-diff} again. 
Using the fact that $\Delta(s)$ scales either as inverse polynomially or inverse exponentially (shown in Ref.~\cite{Somma:2012kx}; see also Fig.~\ref{fig:spectrum}), and the fact that the random variable $\mathcal{N}(0,4)$ has no scaling with problem size we can conclude that, typically, at  first order in perturbation theory, the perturbed problem has an $\mathcal{O}(1)$ gap. (A similar argument establishes that the gap is $\mathcal{O}(1)$  for the case of the LS model as well.)

This argument will only work if $\epsilon \mathcal{N}(0,4)$ does not make the right-hand side of Eq.~\eqref{eq:gappert} negative; if the right-hand side is negative, the perturbative approximation breaks down. The RHS is distributed according to $\mathcal{N}(\Delta(s),4\epsilon^2)$. This is a Gaussian centered around a positive mean, and thus the chance of this distribution sampling negative values is small for small $\epsilon$. So, heuristically, we expect the perturbative argument to work in typical instances. To corroborate the conclusion of this argument, we display the scaling of the gaps with problem size in Fig.~\ref{fig:noisygapsvsn}. It is apparent that the long-range models exhibit a constant scaling with problem size, while the short-range models exhibit an exponential scaling. 

Note that the perturbative argument presented above is not directly applicable for the short-range noise models, because for these models matrix elements that are arbitrarily far apart in the column basis are not normally distributed.

Let us see how the perturbative lifting of the spectral gap helps explain the non-monotonic dependence on noise strength seen in Fig.~\ref{fig:mdnpgsvsepsmanyn}. For $\epsilon \approx 0$, the algorithm succeeds because of the diabatic transitions from the ground state to the first excited state and then back down to the ground state; recall Figs.~\ref{fig:gtqatf250} and~\ref{fig:noiselesspopsn20}. As we increase $\epsilon$ from zero, the slight lifting of the gap interferes with these diabatic transitions, leading to a somewhat smaller success probability. This corresponds to the local minimum in Fig.~\ref{fig:mdnpgsvsepsmanyn}. As we increase $\epsilon$ further, the gap increases more and this causes the dynamics to turn adiabatic, which increases the success probability. This corresponds to the second peak in Fig.~\ref{fig:mdnpgsvsepsmanyn}. These two effects can be seen in Fig.~\ref{fig:popdynamicssymnoise}, which shows typical instances of the noisy spectrum and the noisy dynamics under the LS noise model, at $\epsilon=10^{-3}$ and $\epsilon=10^{-2}$, for $n=20$. Figure~\ref{fig:typicalspectrum} shows that both noise realizations increase the spectral gap from its value in the noiseless case, more so for the higher $\epsilon$ realization. Figure~\ref{fig:10-3dynamics} shows how the diabatic transitions are scrambled due to the noise. Then, as we increase the noise to $\epsilon = 10^{-2}$ in Fig.~\ref{fig:10-2dynamics}, we observe the onset of adiabaticity. 

As we increase $\epsilon$ to values greater than $10^{-2}$, the success probability falls off because even if the dynamics are adiabatic, the noisy spectrum and eigenstates have little relationship with the noiseless spectrum and eigenstates. To corroborate this, in Appendix~\ref{app:pertdecay}, we show using perturbation theory that the overlap between the unperturbed ground state and the perturbed ground state decays as $1- \mathcal{O}(\epsilon^2)$, which suggests that as we increase $\epsilon$, even if the dynamics are adiabatic, the ground state found at the end of the noisy evolution has low overlap with the EXIT vertex.

The perturbative lifting of the gap also explains why the long-range models exhibit an exponential quantum speedup. Indeed, because the noise induces adiabaticity for a certain range of values of $\epsilon$, then as long as the overlap between the unperturbed ground state and the perturbed ground state remains significant, the noisy quantum system can still solve the problem by evolving adiabatically as long as the anneal timescale is greater than the adiabatic timescale. The latter is given by the inverse of the perturbed gap squared, which is $\mathcal{O}(1)$, multiplied the norm of the Hamiltonian, which is $\mathcal{O}(\mathrm{poly}(n))$. The anneal timescale is chosen such that it provides a speedup in the noiseless case [Eq.~\eqref{eqt:tf} represents one such choice], which means the noiseless dynamics are adiabatic relative to the polynomially small gap between the first and second excited states (see Sec.~\ref{sec:gtqadynamics}), and therefore, the long-range noise dynamics will be adiabatic relative the constant gap between the ground and first excited state.

A natural question arises at this stage. Is the anomalous behavior of success probability with increasing system simply due to the perturbation increasing the energy scale of the system? That is, because the long-range noise matrices are selected from the GOE (and variants thereof), the norm of the perturbation increases with system size as $\sqrt{n}$; one might think that since the energies in the system increase, then so do its energy gaps, and then the larger gaps are responsible for the increase in success probability. 

\begin{figure}[!htbp]
\centering
\includegraphics[width=\columnwidth]{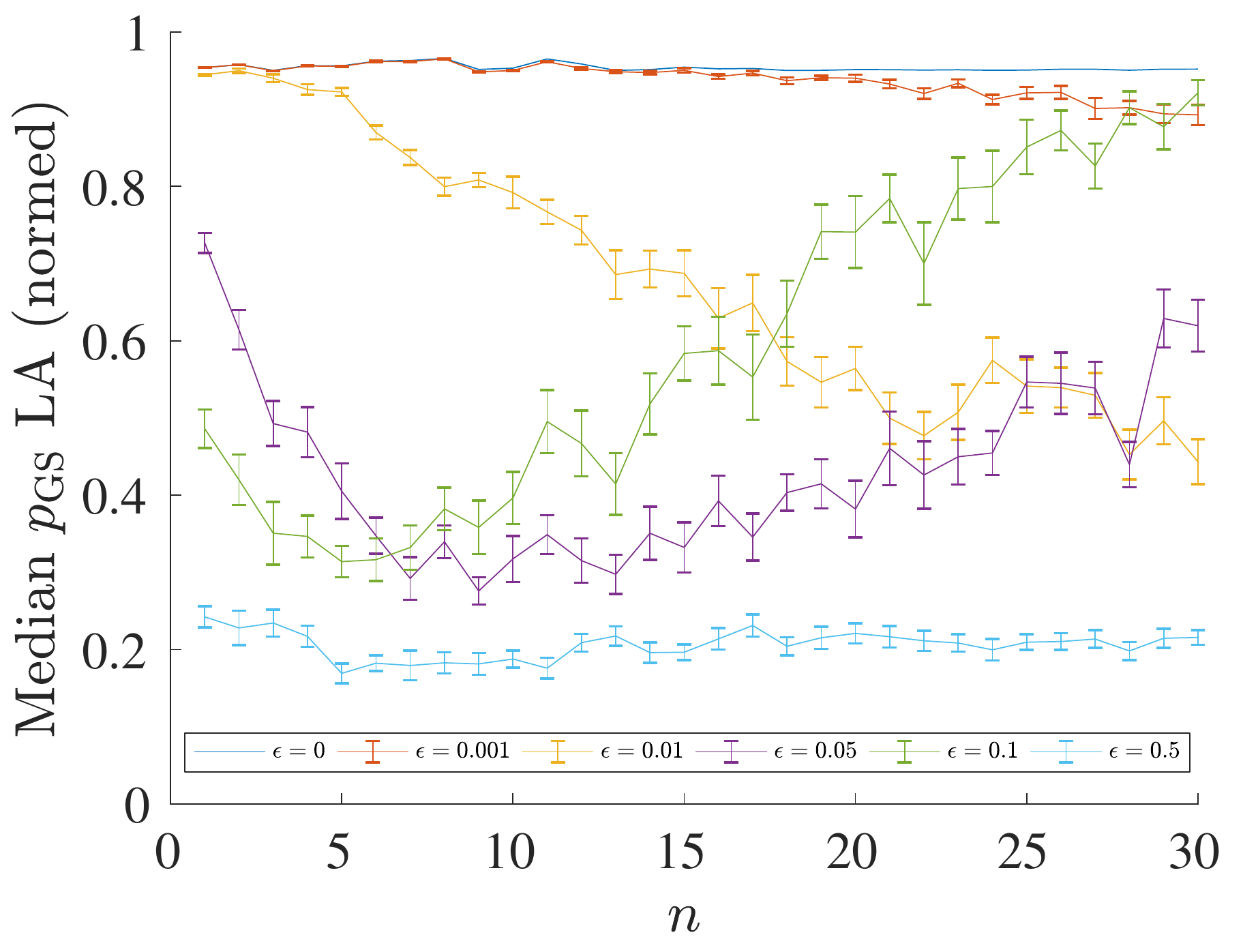}
\caption{(Color online) The median success probability, $p_\mathrm{GS}$, for the \emph{normalized} long-range asymmetric noise case at the end of an evolution of duration $t_f^\mathrm{Th}(n)$, with $\epsilon$ increasing from top to bottom at $n=1$.$t_f^\mathrm{Th}(n)$ is chosen so that the success probability for the noiseless probability is just above $0.95$. Error bars were obtained by bootstrap sampling over $300$ realizations of the noise.}
\label{fig:normLAmdnpgsvsn}
\end{figure}

To test this explanation, we checked what happens after 
normalization of the perturbation matrix by 
its largest eigenvalue. This normalization ensures that the perturbation never adds an amount of energy that scales with system size. If it were true that the perturbative lifting of the gap is due to pumping energy into the system, we would expect the perturbative lifting to disappear upon normalization, and consequently also expect the anomalous behavior of success probability with increasing system size to disappear upon normalization. However, as seen in Fig.~\ref{fig:normLAmdnpgsvsn}, the non-monotonic  dependence of the success probability on $n$ for intermediate $\epsilon$ values continues to hold even after normalization. (This will in turn lead to non-monotonic dependence of the success probability on $\epsilon$ for larger values of $n$, analogous to the behavior seen in Fig.~\ref{fig:mdnpgsvsepsmanyn}.) This is qualitatively similar to the behavior seen in Fig.~\ref{fig:LAmanyeps} for the long-range asymmetric noise model, where the perturbation matrix $h$ is not normalized.

\section{Summary and Conclusions}
\label{sec:conclusion}

We have analyzed the quantum annealing algorithm for the glued trees problem under four different 
noise models: long-range vs. short-range and reflection-symmetric vs. reflection-asymmetric. These are oracular noise models: they add a Gaussian perturbation, of different forms, to the Hamiltonian evolution. We studied the success probability---i.e. the probability finding the EXIT vertex---at the end of the Schr{\"o}dinger evolutions for these different noise models. 

We found that the long-range noise models display a perturbative lifting of the spectral gap which causes the dynamics to transition from diabatic to adiabatic. This allows the algorithm subject to long-range noise models to solve the glued trees problem in polynomial time and hence display an exponential quantum speedup over the noiseless classical algorithm. This seems surprising, since it associates a robustness to noise with a quantum algorithm exhibiting exponential speedup. However, we argue that in fact this speedup is misleading, because it disappears when we compare the quantum algorithm to an appropriate classical analogue. More precisely, a classical random walk that has long-range transition probabilities will also be able to solve the problem in polynomial time and hence display an exponential speedup over the noiseless classical algorithm. 

This analysis highlights that care must be taken in the selection of noise-models for oracular algorithms. Typically, oracles are hard to realize physically, so we must select phenomenological noise models for them. But in so choosing, we might end up with a model that changes the nature of the problem, which is what occurred in the long-range noise models. More precisely, the classical long-range noisy version of the algorithm changed its complexity from exponential into polynomial, so a quantum polynomially scaling algorithm for the problem cannot count as providing an exponential quantum speedup. 

It is instructive to compare this with the results of Ref.~\cite{cross2015quantum}, which analyzed the problem of learning the class of $n$-bit parity functions by making queries to a (noisy) quantum example oracle. There, the quantum algorithm has a linear speedup in the noiseless case, while it has a superpolynomial speedup when both the classical and quantum oracles are noisy. This happens upon depolarizing the qubits at the oracle's output at any constant nonzero rate.

For the glued trees problem we found a weaker result under the symmetric and asymmetric short-range noise models, which retain the exponential complexity of the classical problem. For sufficiently weak oracle noise, the quantum annealing algorithm retains a polynomial quantum speedup over the noiseless classical algorithm. But, for sufficiently strong oracle noise even the polynomial speedup is lost. The fact that for all values of the oracle noise the short-range noise models result in a loss of exponential speedup demonstrates that the exponential speedup of the glued-trees algorithm is not robust to noise. 

We conjecture that more broadly, in the absence of fault tolerant error correction, exponential speedups cannot be obtained in any physical implementation of quantum annealing. This should not necessarily be a cause for pessimism: we are not ruling out polynomial speedups, which remain highly interesting and valuable.

\acknowledgements

We are especially grateful to Hidetoshi Nishimori for an important observation regarding perturbative gap lifting. We also thank Milad Marvian and Evgeny Mozgunov for useful discussions and Huo Chen and Richard Li for advice regarding parallel computation. The research is based upon work (partially) supported by the Office of
the Director of National Intelligence (ODNI), Intelligence Advanced
Research Projects Activity (IARPA), via the U.S. Army Research Office
contract W911NF-17-C-0050. The views and conclusions contained herein are
those of the authors and should not be interpreted as necessarily
representing the official policies or endorsements, either expressed or
implied, of the ODNI, IARPA, or the U.S. Government. The U.S. Government
is authorized to reproduce and distribute reprints for Governmental
purposes notwithstanding any copyright annotation thereon. Computation for the work described in this paper was supported by the University of Southern California's Center for High-Performance Computing (hpc.usc.edu).

\appendix

\section{Hamiltonian in the column basis}
\label{app:colbasis}

Here we show how the matrix elements of the noiseless QA Hamiltonian are obtained in the column basis. I.e., we show how Eqs.~\eqref{eq:colbasisH} are obtained when Eq.~\eqref{eq:QAHam} is written in the basis defined in Eq.~\eqref{eq:colbasisdef}.

$H_0$ and $H_1$ are straightforward: they are represented as $-\ket{\col_0}\bra{\col_0}$ and $-\ket{\col_{2n+1}}\bra{\col_{2n+1}}$ respectively. This immediately yields Eqs.~\eqref{eq:colENT} and~\eqref{eq:colEXIT}. 
Consider the adjacency matrix $A$. Consider first $j < n$. In this case, in the column basis, $A_{j,j+1}$ is
\begin{align} 
\braket{\col_j | A | \col_{j+1}} &= \bra{\col_j} \left( \sum_{(x,x^\pr) \in E}  \ket{x}\bra{x^\pr} \right) \ket{\col_{j+1}} \\
&=\frac{1}{\sqrt{N_j N_{j+1}}}  \sum_{\substack{x \in \col_j ; x^\pr \in \col_{j+1} \\  (x,x^\pr) \in E}} 1 \\
&= \frac{N_{j+1}}{\sqrt{N_j N_{j+1}}} \\
&= \sqrt{\frac{N_{j+1}}{N_j}} = \sqrt{\frac{2^{j+1}}{2^j}} = \sqrt{2},
\end{align}
where we have used the fact that for all columns to the left of the central glue, every vertex has exactly one edge connecting it to the column to its left. We have also used that for $j \leq n$, $N_j = 2^j$.
A parallel calculation will go through for $j > n+1$. Now, for $j=n$:
\begin{align}
\braket{\col_n | A | \col_{n+1}} &= \frac{1}{\sqrt{N_n N_{n+1}}} \times 2 N_{n+1} \\
&= 2 \times \sqrt{\frac{N_{n+1}}{N_n}} = 2,
\end{align}
where we used that, at the central glue, there are two edges to every vertex. We have also used the fact that the number of vertices in the two columns at the glue is equal. Thus, we have:
\beq
\braket{\col_j | A | \col_{j+1}} = \begin{cases} 2, & j=n \\ \sqrt{2}, & o.w. \end{cases}.
\eeq
For convenience, we redefine $A$ such that $A_\mathrm{new} \equiv A_\mathrm{old}/\sqrt{2}$, giving us:
\beq\label{eq:adjcol}
\braket{\col_j | A | \col_{j+1}} = \begin{cases} \sqrt{2}, & j=n \\ 1, & o.w. \end{cases},
\eeq
which yields Eqs.~\eqref{eq:coladj} and~\eqref{eq:colglue}.

\section{Qubit versions of small glued trees instances}\label{app:qubitgt}

If we wanted to implement the glued-trees algorithm in a qubit system, we would need nonlocal and difficult-to-engineer interactions. We do not provide a mathematical proof of this claim, but the basic point can be illustrated using the case of $n=1$, the smallest instance of the glued-trees problem which has $6$ vertices. Consider the shortest possible naming system, in which each vertex is labelled by a length-3 bit-string. (Set aside, for the moment, the concern about this naming system interfering with the proof of classical hardness.) This means we can implement the QA Hamiltonian using $3$ qubits. 

\begin{figure}[!h]
\centering
\includegraphics[width=\columnwidth]{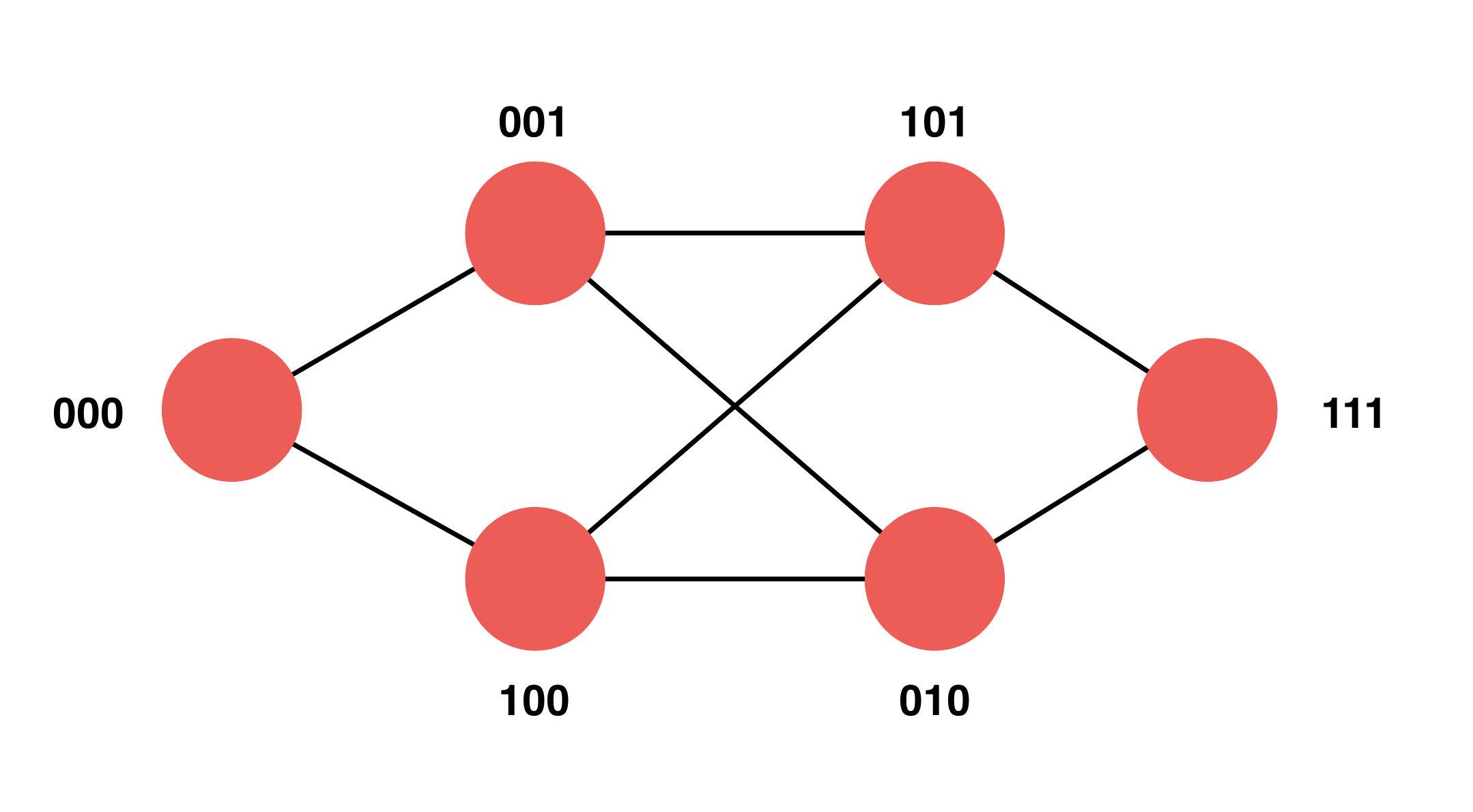}
\caption{(Color online) The smallest instance of the glued-trees problem. The graph is labelled with $3$ bits. The naming is chosen so as to minimize the Hamming distance between vertices joined by an edge.}
\label{fig:smalltree}
\end{figure}

We will name the vertices such that the Hamming distance between two vertices connected by an edge on the graph is as small as possible. This is done in an attempt to reduce the need for many-body terms as much as possible. For the case of $n=1$, this is done as shown in Fig.~\ref{fig:smalltree}.

Clearly the adjacency matrix $A$ of this graph is
\begin{align}
A = &\ket{v_0}\bra{v_1}+\ket{v_0}\bra{v_2}+\ket{v_1}\bra{v_3}+\ket{v_1}\bra{v_4}+\ket{v_2}\bra{v_3} \nonumber\\ 
&+\ket{v_2}\bra{v_4}+\ket{v_3}\bra{v_5}+\ket{v_4}\bra{v_5} + h.c.,
\end{align}
where $v_0 \equiv 000$, $v_1 \equiv 001$, $v_2 \equiv 100$, $v_3 \equiv 101$, $v_4 \equiv 010$, and $v_5 \equiv 111$.

Let us convert this adjacency matrix into a local Hamiltonian. Let $\{I_i, X_i, Y_i, Z_i\}$ denote the 
Pauli operators on the $i$-th qubit ($i=1,2,3$). We can compute $\Tr[A X_1]$, $\Tr[A X_1X_2]$,  $\Tr[A X_1Y_2]$, and so on, for all the $4^3 = 64$ terms to obtain the coefficients of these terms in the Pauli representation of $A$. Doing this, we get
\begin{align}
A = &\frac{1}{4}\left( 2 X_1 + X_2 + 2 X_3 + X_1 X_2  + X_2 X_3 \right. \nonumber \\
&+ X_1 X_3 + Y_1 Y_2 + Y_2 Y_3  \nonumber \\
& - Y_1 Y_3 + 2 X_1 Z_2 - X_2 Z_3 - Z_1 X_2 \nonumber \\
&+ 2 Z_2 Y_3 - X_1 Z_2 X_3 + X_1 X_2 Z_3  \nonumber\\
& + Y_1 Y_2 Z_3 + Y_1 Z_3 Y_3 + Z_1 X_2 X_3 \nonumber \\
&\left. + Z_1 X_2 Z_3 + Z_1 Y_2 Y_3 \right).
\end{align}
Note that there are no $Z$ terms because $A$ does not contain any diagonal matrix elements. We can already see that in order to implement this adjacency matrix we need $3$-body interactions, and also ``cross-term" interactions, i.e., interactions that couple, e.g., $Z$ and $X$. 

To complete the analysis, let us similarly write the Pauli representations of $H_0$ and $H_1$. This yields
\bes
\begin{align}
H_0 = &\frac{1}{8}\left(Z_1 + Z_2 + Z_3 + Z_1 Z_2 \right. \\ \nonumber
&\left. + Z_2 Z_3 + Z_1 Z_3 + Z_1 Z_2 Z_3 \right)\\
H_1 = &\frac{1}{8}\left(-Z_1 - Z_2 - Z_3 + Z_1 Z_2 + Z_2 Z_3 \right. \\ \nonumber
& \left. + Z_1 Z_3 - Z_1 Z_2 Z_3 \right).
\end{align}
\ees
Again $3$-body interactions are required.

We have not ruled out the possibility that there exists an easily computed naming system for a general instance of the glued-trees problem, which generates a Hamiltonian representation for the QA Hamiltonian in such a way that the representation has small and constant locality, and the interactions required are simple (such as $XX$ or $YY$). However, we are not aware of an explicit method for doing so and conjecture that none exists.

\section{The Gaussian Orthogonal Ensemble}
\label{app:GOE}

The GOE is the measure over the set of $N \times N$ real symmetric matrices described by
\beq \label{eq:goe}
\mathrm{Pr}(h)dh = c_N \exp\left[ - \frac{1}{2\sigma^2} \Tr(h^2) \right] \prod_i d h_{ii} \prod_{i<j} d h_{ij},
\eeq
where $\sigma$ is the so-called scale factor, $c_N$ is a normalization, $d h_{ii}$ and $d h_{ij}$ are the standard Lebesgue measure. If $\{ \lambda_i \}$ are the eigenvalues of $h$, then $\Tr(h^2) = \sum_i \lambda_i^2$. Therefore, the peak of this distribution is centered around matrices with eigenvalues close to $0$. The GOE is invariant under conjugation by orthogonal matrices: i.e., if Eq.~\eqref{eq:goe} holds in one basis, then it also holds in another basis related by an orthogonal transformation to the first basis. See, e.g., Ref.~\cite{mehta2004random} for more details.\\

\section{Perturbative decay of overlap between the noisy and noiseless ground states}\label{app:pertdecay}

Here we calculate how the overlap between the noisy and noiseless groundstates decays as a function of $\epsilon$ for the case of the long-range noise models. This is done using perturbation theory, so we expect it to be valid for small $\epsilon$. 

Consider first the perturbative expansion for the perturbed ground state.
\beq
\ket{\tilde{\phi}_0} = c(\epsilon)  \ket{\phi_0} + \epsilon \sum_{k > 0} \ket{\phi_k} \frac{h_{k0}}{E_0 - E_k} + \mathcal{O}(\epsilon^2),
\eeq
Since $c(\epsilon) = \braket{\phi_0 | \tilde{\phi}_0 }$, this is the quantity we care about, namely, the overlap between the noisy and the noiseless ground states. Imposing $\braket{\tilde{\phi}_0 | \tilde{\phi}_0} = 1$ (we are free to choose the normalization; see, e.g., Chapter 5 of Ref.~\cite{sakurai2017modern}) and assuming $c(\epsilon) \in \mathbb{R}$, we get
\bes
\begin{align}
c(\epsilon) &\approx \left[ 1 - \epsilon^2 \sum_{k>0}  \frac{h_{k0}^2}{(E_0 - E_k)^2} \right]^{\frac{1}{2}} \\
&\approx 1 - \frac{\epsilon^2}{2} \sum_{k>0}  \frac{h_{k0}^2}{(E_0 - E_k)^2} .\label{eq:cepsilon}
\end{align}
\ees
Thus the fidelity between the noisy and noiseless ground states goes as $1- \mathcal{O}(\epsilon^2)$.

\bibliographystyle{apsrev4-1}
\bibliography{refs}

\begin{thebibliography}{25}%
\makeatletter
\providecommand \@ifxundefined [1]{%
 \@ifx{#1\undefined}
}%
\providecommand \@ifnum [1]{%
 \ifnum #1\expandafter \@firstoftwo
 \else \expandafter \@secondoftwo
 \fi
}%
\providecommand \@ifx [1]{%
 \ifx #1\expandafter \@firstoftwo
 \else \expandafter \@secondoftwo
 \fi
}%
\providecommand \natexlab [1]{#1}%
\providecommand \enquote  [1]{``#1''}%
\providecommand \bibnamefont  [1]{#1}%
\providecommand \bibfnamefont [1]{#1}%
\providecommand \citenamefont [1]{#1}%
\providecommand \href@noop [0]{\@secondoftwo}%
\providecommand \href [0]{\begingroup \@sanitize@url \@href}%
\providecommand \@href[1]{\@@startlink{#1}\@@href}%
\providecommand \@@href[1]{\endgroup#1\@@endlink}%
\providecommand \@sanitize@url [0]{\catcode `\\12\catcode `\$12\catcode
  `\&12\catcode `\#12\catcode `\^12\catcode `\_12\catcode `\%12\relax}%
\providecommand \@@startlink[1]{}%
\providecommand \@@endlink[0]{}%
\providecommand \url  [0]{\begingroup\@sanitize@url \@url }%
\providecommand \@url [1]{\endgroup\@href {#1}{\urlprefix }}%
\providecommand \urlprefix  [0]{URL }%
\providecommand \Eprint [0]{\href }%
\providecommand \doibase [0]{http://dx.doi.org/}%
\providecommand \selectlanguage [0]{\@gobble}%
\providecommand \bibinfo  [0]{\@secondoftwo}%
\providecommand \bibfield  [0]{\@secondoftwo}%
\providecommand \translation [1]{[#1]}%
\providecommand \BibitemOpen [0]{}%
\providecommand \bibitemStop [0]{}%
\providecommand \bibitemNoStop [0]{.\EOS\space}%
\providecommand \EOS [0]{\spacefactor3000\relax}%
\providecommand \BibitemShut  [1]{\csname bibitem#1\endcsname}%
\let\auto@bib@innerbib\@empty
\bibitem [{\citenamefont {Kadowaki}\ and\ \citenamefont
  {Nishimori}(1998)}]{kadowaki_quantum_1998}%
  \BibitemOpen
  \bibfield  {author} {\bibinfo {author} {\bibfnamefont {T.}~\bibnamefont
  {Kadowaki}}\ and\ \bibinfo {author} {\bibfnamefont {H.}~\bibnamefont
  {Nishimori}},\ }\href
  {http://journals.aps.org/pre/abstract/10.1103/PhysRevE.58.5355} {\bibfield
  {journal} {\bibinfo  {journal} {Phys. Rev. E}\ }\textbf {\bibinfo {volume}
  {58}},\ \bibinfo {pages} {5355} (\bibinfo {year} {1998})}\BibitemShut
  {NoStop}%
\bibitem [{\citenamefont {Farhi}\ \emph {et~al.}(2001)\citenamefont {Farhi},
  \citenamefont {Goldstone}, \citenamefont {Gutmann}, \citenamefont {Lapan},
  \citenamefont {Lundgren},\ and\ \citenamefont {Preda}}]{farhi_quantum_2001}%
  \BibitemOpen
  \bibfield  {author} {\bibinfo {author} {\bibfnamefont {E.}~\bibnamefont
  {Farhi}}, \bibinfo {author} {\bibfnamefont {J.}~\bibnamefont {Goldstone}},
  \bibinfo {author} {\bibfnamefont {S.}~\bibnamefont {Gutmann}}, \bibinfo
  {author} {\bibfnamefont {J.}~\bibnamefont {Lapan}}, \bibinfo {author}
  {\bibfnamefont {A.}~\bibnamefont {Lundgren}}, \ and\ \bibinfo {author}
  {\bibfnamefont {D.}~\bibnamefont {Preda}},\ }\href {\doibase
  10.1126/science.1057726} {\bibfield  {journal} {\bibinfo  {journal}
  {Science}\ }\textbf {\bibinfo {volume} {292}},\ \bibinfo {pages} {472}
  (\bibinfo {year} {2001})}\BibitemShut {NoStop}%
\bibitem [{\citenamefont {Finnila}\ \emph {et~al.}(1994)\citenamefont
  {Finnila}, \citenamefont {Gomez}, \citenamefont {Sebenik}, \citenamefont
  {Stenson},\ and\ \citenamefont {Doll}}]{finnila_quantum_1994}%
  \BibitemOpen
  \bibfield  {author} {\bibinfo {author} {\bibfnamefont {A.~B.}\ \bibnamefont
  {Finnila}}, \bibinfo {author} {\bibfnamefont {M.~A.}\ \bibnamefont {Gomez}},
  \bibinfo {author} {\bibfnamefont {C.}~\bibnamefont {Sebenik}}, \bibinfo
  {author} {\bibfnamefont {C.}~\bibnamefont {Stenson}}, \ and\ \bibinfo
  {author} {\bibfnamefont {J.~D.}\ \bibnamefont {Doll}},\ }\href {\doibase
  10.1016/0009-2614(94)00117-0} {\bibfield  {journal} {\bibinfo  {journal}
  {Chemical Physics Letters}\ }\textbf {\bibinfo {volume} {219}},\ \bibinfo
  {pages} {343} (\bibinfo {year} {1994})}\BibitemShut {NoStop}%
\bibitem [{\citenamefont {Brooke}\ \emph {et~al.}(1999)\citenamefont {Brooke},
  \citenamefont {Bitko}, \citenamefont {F.}, \citenamefont {Rosenbaum},\ and\
  \citenamefont {Aeppli}}]{Brooke1999}%
  \BibitemOpen
  \bibfield  {author} {\bibinfo {author} {\bibfnamefont {J.}~\bibnamefont
  {Brooke}}, \bibinfo {author} {\bibfnamefont {D.}~\bibnamefont {Bitko}},
  \bibinfo {author} {\bibfnamefont {T.}~\bibnamefont {F.}}, \bibinfo {author}
  {\bibnamefont {Rosenbaum}}, \ and\ \bibinfo {author} {\bibfnamefont
  {G.}~\bibnamefont {Aeppli}},\ }\href {\doibase 10.1126/science.284.5415.779}
  {\bibfield  {journal} {\bibinfo  {journal} {Science}\ }\textbf {\bibinfo
  {volume} {284}},\ \bibinfo {pages} {779} (\bibinfo {year}
  {1999})}\BibitemShut {NoStop}%
\bibitem [{\citenamefont {Santoro}\ \emph {et~al.}(2002)\citenamefont
  {Santoro}, \citenamefont {Marto\v{n}\'{a}k}, \citenamefont {Tosatti},\ and\
  \citenamefont {Car}}]{Santoro}%
  \BibitemOpen
  \bibfield  {author} {\bibinfo {author} {\bibfnamefont {G.~E.}\ \bibnamefont
  {Santoro}}, \bibinfo {author} {\bibfnamefont {R.}~\bibnamefont
  {Marto\v{n}\'{a}k}}, \bibinfo {author} {\bibfnamefont {E.}~\bibnamefont
  {Tosatti}}, \ and\ \bibinfo {author} {\bibfnamefont {R.}~\bibnamefont
  {Car}},\ }\href {http://science.sciencemag.org/content/295/5564/2427}
  {\bibfield  {journal} {\bibinfo  {journal} {Science}\ }\textbf {\bibinfo
  {volume} {295}},\ \bibinfo {pages} {2427} (\bibinfo {year}
  {2002})}\BibitemShut {NoStop}%
\bibitem [{\citenamefont {Albash}\ and\ \citenamefont
  {Lidar}(2018)}]{Albash-Lidar:RMP}%
  \BibitemOpen
  \bibfield  {author} {\bibinfo {author} {\bibfnamefont {T.}~\bibnamefont
  {Albash}}\ and\ \bibinfo {author} {\bibfnamefont {D.~A.}\ \bibnamefont
  {Lidar}},\ }\href {https://link.aps.org/doi/10.1103/RevModPhys.90.015002}
  {\bibfield  {journal} {\bibinfo  {journal} {Reviews of Modern Physics}\
  }\textbf {\bibinfo {volume} {90}},\ \bibinfo {pages} {015002} (\bibinfo
  {year} {2018})}\BibitemShut {NoStop}%
\bibitem [{\citenamefont {Jansen}\ \emph {et~al.}(2007)\citenamefont {Jansen},
  \citenamefont {Ruskai},\ and\ \citenamefont {Seiler}}]{Jansen:07}%
  \BibitemOpen
  \bibfield  {author} {\bibinfo {author} {\bibfnamefont {S.}~\bibnamefont
  {Jansen}}, \bibinfo {author} {\bibfnamefont {M.-B.}\ \bibnamefont {Ruskai}},
  \ and\ \bibinfo {author} {\bibfnamefont {R.}~\bibnamefont {Seiler}},\ }\href
  {http://scitation.aip.org/content/aip/journal/jmp/48/10/10.1063/1.2798382}
  {\bibfield  {journal} {\bibinfo  {journal} {J. Math. Phys.}\ }\textbf
  {\bibinfo {volume} {48}},\ \bibinfo {pages} {102111} (\bibinfo {year}
  {2007})}\BibitemShut {NoStop}%
\bibitem [{\citenamefont {Das}\ and\ \citenamefont
  {Chakrabarti}(2008)}]{RevModPhys.80.1061}%
  \BibitemOpen
  \bibfield  {author} {\bibinfo {author} {\bibfnamefont {A.}~\bibnamefont
  {Das}}\ and\ \bibinfo {author} {\bibfnamefont {B.~K.}\ \bibnamefont
  {Chakrabarti}},\ }\href {\doibase 10.1103/RevModPhys.80.1061} {\bibfield
  {journal} {\bibinfo  {journal} {Rev. Mod. Phys.}\ }\textbf {\bibinfo {volume}
  {80}},\ \bibinfo {pages} {1061} (\bibinfo {year} {2008})}\BibitemShut
  {NoStop}%
\bibitem [{\citenamefont {Childs}\ \emph {et~al.}(2003)\citenamefont {Childs},
  \citenamefont {Cleve}, \citenamefont {Deotto}, \citenamefont {Farhi},
  \citenamefont {Gutmann},\ and\ \citenamefont
  {Spielman}}]{childs2003exponential}%
  \BibitemOpen
  \bibfield  {author} {\bibinfo {author} {\bibfnamefont {A.~M.}\ \bibnamefont
  {Childs}}, \bibinfo {author} {\bibfnamefont {R.}~\bibnamefont {Cleve}},
  \bibinfo {author} {\bibfnamefont {E.}~\bibnamefont {Deotto}}, \bibinfo
  {author} {\bibfnamefont {E.}~\bibnamefont {Farhi}}, \bibinfo {author}
  {\bibfnamefont {S.}~\bibnamefont {Gutmann}}, \ and\ \bibinfo {author}
  {\bibfnamefont {D.~A.}\ \bibnamefont {Spielman}},\ }in\ \href {\doibase
  10.1145/780542.780552} {\emph {\bibinfo {booktitle} {Proceedings of the
  thirty-fifth annual ACM symposium on Theory of computing}}}\ (\bibinfo
  {organization} {ACM},\ \bibinfo {year} {2003})\ pp.\ \bibinfo {pages}
  {59--68}\BibitemShut {NoStop}%
\bibitem [{\citenamefont {Somma}\ \emph {et~al.}(2012)\citenamefont {Somma},
  \citenamefont {Nagaj},\ and\ \citenamefont {Kieferov{\'a}}}]{Somma:2012kx}%
  \BibitemOpen
  \bibfield  {author} {\bibinfo {author} {\bibfnamefont {R.~D.}\ \bibnamefont
  {Somma}}, \bibinfo {author} {\bibfnamefont {D.}~\bibnamefont {Nagaj}}, \ and\
  \bibinfo {author} {\bibfnamefont {M.}~\bibnamefont {Kieferov{\'a}}},\ }\href
  {http://link.aps.org/doi/10.1103/PhysRevLett.109.050501} {\bibfield
  {journal} {\bibinfo  {journal} {Phys. Rev. Lett.}\ }\textbf {\bibinfo
  {volume} {109}},\ \bibinfo {pages} {050501} (\bibinfo {year}
  {2012})}\BibitemShut {NoStop}%
\bibitem [{\citenamefont {{D.R. Simon}}(1994)}]{Simon:94}%
  \BibitemOpen
  \bibfield  {author} {\bibinfo {author} {\bibnamefont {{D.R. Simon}}},\ }in\
  \href {https://doi.org/10.1109/SFCS.1994.365701} {\emph {\bibinfo {booktitle}
  {{Proceedings of the 35th Annual Symposium on the Foundations of Computer
  Science}}}},\ \bibinfo {editor} {edited by\ \bibinfo {editor} {\bibnamefont
  {{S. Goldwasser}}}}\ (\bibinfo  {publisher} {{IEEE Computer Society}},\
  \bibinfo {address} {{Los Alamitos, CA}},\ \bibinfo {year} {1994})\ p.\
  \bibinfo {pages} {116}\BibitemShut {NoStop}%
\bibitem [{\citenamefont {Shor}(1994)}]{Shor:94}%
  \BibitemOpen
  \bibfield  {author} {\bibinfo {author} {\bibfnamefont {P.~W.}\ \bibnamefont
  {Shor}},\ }\bibfield  {booktitle} {\emph {\bibinfo {booktitle} {Foundations
  of Computer Science, 1994 Proceedings., 35th Annual Symposium on}},\ }\href
  {http://ieeexplore.ieee.org/document/365700/} {\bibfield  {journal} {\bibinfo
   {journal} {35th Annual Symposium on Foundations of Computer Science, 1994
  Proceedings}\ ,\ \bibinfo {pages} {124}} (\bibinfo {year} {20-22 Nov
  1994})}\BibitemShut {NoStop}%
\bibitem [{\citenamefont {Breuer}\ and\ \citenamefont
  {Petruccione}(2002)}]{Breuer:2002}%
  \BibitemOpen
  \bibfield  {author} {\bibinfo {author} {\bibfnamefont {H.-P.}\ \bibnamefont
  {Breuer}}\ and\ \bibinfo {author} {\bibfnamefont {F.}~\bibnamefont
  {Petruccione}},\ }\href
  {https://global.oup.com/academic/product/the-theory-of-open-quantum-systems-9780198520634?cc=us&lang=en&}
  {\emph {\bibinfo {title} {The Theory of Open Quantum Systems}}}\ (\bibinfo
  {publisher} {Oxford University Press},\ \bibinfo {year} {2002})\BibitemShut
  {NoStop}%
\bibitem [{Note1()}]{Note1}%
  \BibitemOpen
  \bibinfo {note} {An algorithm $A$ has an exponential (polynomial) speedup
  over another algorithm $B$ if the asymptotic scaling of $A$ is an exponential
  (polynomial) function of the asymptotic scaling of $B$.}\BibitemShut {Stop}%
\bibitem [{Note2()}]{Note2}%
  \BibitemOpen
  \bibinfo {note} {That all the vertices, except the ENTRANCE and the EXIT
  vertices, have equal degree is crucial to avoid the easy solution of this
  problem by a backtracking classical random walk. See Ref.~\cite
  {childs2003exponential}.}\BibitemShut {Stop}%
\bibitem [{Note3()}]{Note3}%
  \BibitemOpen
  \bibinfo {note} {The labeling scheme used for the vertices does not affect
  the performance of the quantum algorithm.}\BibitemShut {Stop}%
\bibitem [{\citenamefont {{N. Shenvi, K.R. Brown, and K.B.
  Whaley}}(2003)}]{Shenvi:03}%
  \BibitemOpen
  \bibfield  {author} {\bibinfo {author} {\bibnamefont {{N. Shenvi, K.R. Brown,
  and K.B. Whaley}}},\ }\href {https://doi.org/10.1103/PhysRevA.68.052313}
  {\bibfield  {journal} {\bibinfo  {journal} {Phys. Rev. A}\ }\textbf {\bibinfo
  {volume} {68}},\ \bibinfo {pages} {052313} (\bibinfo {year}
  {2003})}\BibitemShut {NoStop}%
\bibitem [{\citenamefont {Temme}(2014)}]{temme2014runtime}%
  \BibitemOpen
  \bibfield  {author} {\bibinfo {author} {\bibfnamefont {K.}~\bibnamefont
  {Temme}},\ }\href {\doibase 10.1103/PhysRevA.90.022310} {\bibfield  {journal}
  {\bibinfo  {journal} {Physical Review A}\ }\textbf {\bibinfo {volume} {90}},\
  \bibinfo {pages} {022310} (\bibinfo {year} {2014})}\BibitemShut {NoStop}%
\bibitem [{\citenamefont {Cross}\ \emph {et~al.}(2015)\citenamefont {Cross},
  \citenamefont {Smith},\ and\ \citenamefont {Smolin}}]{cross2015quantum}%
  \BibitemOpen
  \bibfield  {author} {\bibinfo {author} {\bibfnamefont {A.~W.}\ \bibnamefont
  {Cross}}, \bibinfo {author} {\bibfnamefont {G.}~\bibnamefont {Smith}}, \ and\
  \bibinfo {author} {\bibfnamefont {J.~A.}\ \bibnamefont {Smolin}},\ }\href
  {\doibase 10.1103/PhysRevA.92.012327} {\bibfield  {journal} {\bibinfo
  {journal} {Physical Review A}\ }\textbf {\bibinfo {volume} {92}},\ \bibinfo
  {pages} {012327} (\bibinfo {year} {2015})}\BibitemShut {NoStop}%
\bibitem [{\citenamefont {Kendon}(2007)}]{kendon2007decoherence}%
  \BibitemOpen
  \bibfield  {author} {\bibinfo {author} {\bibfnamefont {V.}~\bibnamefont
  {Kendon}},\ }\href {\doibase 10.1017/S0960129507006354} {\bibfield  {journal}
  {\bibinfo  {journal} {Mathematical Structures in Computer Science}\ }\textbf
  {\bibinfo {volume} {17}},\ \bibinfo {pages} {1169} (\bibinfo {year}
  {2007})}\BibitemShut {NoStop}%
\bibitem [{\citenamefont {Lockhart}\ \emph {et~al.}(2014)\citenamefont
  {Lockhart}, \citenamefont {Di~Franco},\ and\ \citenamefont
  {Paternostro}}]{lockhart2014glued}%
  \BibitemOpen
  \bibfield  {author} {\bibinfo {author} {\bibfnamefont {J.}~\bibnamefont
  {Lockhart}}, \bibinfo {author} {\bibfnamefont {C.}~\bibnamefont {Di~Franco}},
  \ and\ \bibinfo {author} {\bibfnamefont {M.}~\bibnamefont {Paternostro}},\
  }\href {\doibase https://doi.org/10.1016/j.physleta.2013.11.034} {\bibfield
  {journal} {\bibinfo  {journal} {Physics Letters A}\ }\textbf {\bibinfo
  {volume} {378}},\ \bibinfo {pages} {338} (\bibinfo {year}
  {2014})}\BibitemShut {NoStop}%
\bibitem [{\citenamefont {Roland}\ and\ \citenamefont
  {Cerf}(2005)}]{PhysRevA.71.032330}%
  \BibitemOpen
  \bibfield  {author} {\bibinfo {author} {\bibfnamefont {J.}~\bibnamefont
  {Roland}}\ and\ \bibinfo {author} {\bibfnamefont {N.~J.}\ \bibnamefont
  {Cerf}},\ }\href {\doibase 10.1103/PhysRevA.71.032330} {\bibfield  {journal}
  {\bibinfo  {journal} {Phys. Rev. A}\ }\textbf {\bibinfo {volume} {71}},\
  \bibinfo {pages} {032330} (\bibinfo {year} {2005})}\BibitemShut {NoStop}%
\bibitem [{\citenamefont {Roland}\ and\ \citenamefont
  {Cerf}(2002)}]{Roland:2002ul}%
  \BibitemOpen
  \bibfield  {author} {\bibinfo {author} {\bibfnamefont {J.}~\bibnamefont
  {Roland}}\ and\ \bibinfo {author} {\bibfnamefont {N.~J.}\ \bibnamefont
  {Cerf}},\ }\href {http://link.aps.org/doi/10.1103/PhysRevA.65.042308}
  {\bibfield  {journal} {\bibinfo  {journal} {Phys. Rev. A}\ }\textbf {\bibinfo
  {volume} {65}},\ \bibinfo {pages} {042308} (\bibinfo {year}
  {2002})}\BibitemShut {NoStop}%
\bibitem [{\citenamefont {Mehta}(2004)}]{mehta2004random}%
  \BibitemOpen
  \bibfield  {author} {\bibinfo {author} {\bibfnamefont {M.~L.}\ \bibnamefont
  {Mehta}},\ }\href
  {https://www.elsevier.com/books/random-matrices/lal-mehta/978-0-12-088409-4}
  {\emph {\bibinfo {title} {Random matrices}}},\ \bibinfo {edition} {3rd}\
  ed.,\ Vol.\ \bibinfo {volume} {142}\ (\bibinfo  {publisher} {Elsevier},\
  \bibinfo {year} {2004})\BibitemShut {NoStop}%
\bibitem [{\citenamefont {Sakurai}\ and\ \citenamefont
  {Napolitano}(2017)}]{sakurai2017modern}%
  \BibitemOpen
  \bibfield  {author} {\bibinfo {author} {\bibfnamefont {J.~J.}\ \bibnamefont
  {Sakurai}}\ and\ \bibinfo {author} {\bibfnamefont {J.}~\bibnamefont
  {Napolitano}},\ }\href {\doibase 10.1017/9781108499996} {\emph {\bibinfo
  {title} {Modern quantum mechanics}}}\ (\bibinfo  {publisher} {Cambridge
  University Press},\ \bibinfo {year} {2017})\BibitemShut {NoStop}%
\end{thebibliography}%

\end{document}